\documentclass[journal]{IEEEtran}
\usepackage{cite}
\usepackage{amsmath,amssymb,amsfonts}
\usepackage{algorithmic}
\usepackage{graphicx}
\usepackage{textcomp}
\usepackage{comment}
\usepackage{array}
\usepackage{multirow}
\usepackage{hyperref}
\usepackage{soul}
\usepackage{xcolor}
\usepackage{subfig}
\usepackage[mode=buildnew]{standalone}

\begin{document}
\title{Has The Physical Layer Matured?}
\author{Mansoor Shafi, \IEEEmembership{Life Fellow, IEEE}, Changlong Xu, Xingqin Lin, Gilwon Lee, Feifei Sun, 
Eko Onggosanusi, 
Oskari Tervo, and Joonyoung Cho

\thanks{Mansoor Shafi is with Spark, New Zealand (e-mail: mansoor.shafi@spark.co.nz).}
\thanks{Changlong Xu is with Qualcomm, China (e-mail: changlon@qti.qualcomm.com).}
\thanks{Xingqin Lin is with NVIDIA, USA (e-mail: xingqinl@nvidia.com).}
\thanks{Gilwon Lee is with Samsung, USA (e-mail: gilwon.lee@samsung.com ).}
\thanks{Feifei Sun is with Samsung, China (e-mail: feifei.sun@samsung.com).}
\thanks{Eko Onggosanusi is with Samsung, USA (e-mail: eko.o@samsung.com).}
\thanks {Oskari Tervo is with Nokia, Finland (e-mail: oskari.tervo@nokia.com).}
\thanks{Joonyoung Cho is with Samsung, USA (e-mail: joonyoung.cho@samsung.com).}
}

\maketitle

\begin{abstract}
The wireless physical (PHY) layer has enabled successive generations of cellular systems through advances in modulation, coding, waveforms, and multiple-input multiple-output (MIMO) transmission. This article assesses whether these techniques are now approaching maturity and where substantial further gains remain possible. Field measurements and quantitative evaluations indicate that many link-level refinements, including constellation shaping, channel-code evolution, reduced-complexity receivers, and waveform enhancements, remain valuable but typically provide bounded gains that must be balanced against implementation complexity and overhead. In contrast, massive MIMO and distributed MIMO offer a more scalable system-level opportunity by increasing the number and quality of usable spatial channels. Their effectiveness relies on time-division duplex reciprocity for scalable channel state information (CSI) acquisition, while practical limitations include calibration, pilot reuse, channel aging, weak pilot reception from cell-edge users, and the fronthaul and synchronization requirements of coherent distributed operation. Artificial intelligence and machine learning (AI/ML) provide complementary opportunities for further PHY layer innovations. The PHY layer is therefore not dead; its most consequential advances will come from deployable spatial processing and CSI acquisition, complemented by targeted link-level and AI/ML-based refinements.
\end{abstract}


\IEEEpeerreviewmaketitle

\section{Introduction}

The wireless physical (PHY) layer remains a rich source of research and innovation. Cellular communication has become pervasive and essential to daily life, while each new generation of cellular systems has raised expectations for capacity, coverage, reliability, and efficiency. As the standardization of sixth-generation (6G) cellular systems progresses \cite{shafi2025industrialviewpointsrantechnologies,10054381,10634051, 11186253}, it is timely to ask which PHY-layer mechanisms can still deliver meaningful improvements in the key performance indicators (KPIs) of future networks. The question is not simply whether further gains are possible, but whether they are scalable and deployable, or instead come with diminishing returns and disproportionate complexity.

Approximately 15 years ago, \cite{5741160} raised the provocative question, \textit{Is the PHY layer dead?} Since then, fourth-generation (4G) systems have matured, fifth-generation (5G) systems have been widely deployed, and 6G standardization has begun \cite{6GWS-25243,shafi2025industrialviewpointsrantechnologies}. The answer is necessarily nuanced. Importantly, the emerging 6G PHY baseline is largely an evolution of 5G: the baseline agreements reuse mature 5G designs for waveforms, modulation, and channel coding, while extending channel bandwidth support to 400~MHz at around 7 GHz \cite{3gpp387601}. This paper therefore examines where further PHY evolution beyond this largely 5G-derived baseline can deliver material and scalable gains. Several classical link-level techniques are already highly optimized and can provide only bounded improvements under practical constraints. In contrast, the spatial domain remains a major source of scalable PHY-layer gains. In particular, massive multiple-input multiple-output (MIMO) and distributed MIMO can increase the number and quality of simultaneously usable spatial channels, offering a fundamentally different opportunity from refining an individual link by a fraction of a decibel \cite{5595728, cellfree}.

The familiar link-level levers nevertheless remain important. High-order quadrature amplitude modulation (QAM) improves peak spectral efficiency, and the 3rd generation partnership project (3GPP) has begun studying modulation and joint channel coding and modulation for 6G \cite{qcom}. However, operation beyond 1024-QAM in downlink and 256-QAM in uplink requires increasingly demanding signal-to-noise ratio (SNR), radio-frequency (RF), and error vector magnitude (EVM) performance. Uniform QAM has an asymptotic shaping gap of up to 1.53~dB under ideal assumptions \cite{qureshi}. Probabilistic shaping (PS) changes the probability of transmitting conventional in-phase/quadrature (I/Q) constellation points, whereas geometric shaping (GS) changes their locations; both can recover part of this gap \cite{pasupathy,shaping,gcs}. Similarly, low-density parity-check (LDPC) codes replaced turbo codes in 5G data channels and are also used in standards such as IEEE 802.11 and digital video broadcast (DVB). LDPC codes provide strong coding gains, high-throughput decoding, and low error floors \cite{turbo,Richardson,Richardson2}. Reduced-complexity receiver architectures are also increasingly important as modulation order and the number of spatial streams grow. Orthogonal frequency-division multiplexing (OFDM) will remain the baseline waveform because of its maturity, implementation efficiency, and backward-compatibility advantages, although waveform refinements may offer benefits in selected scenarios \cite{sixchallenges,shafi2025industrialviewpointsrantechnologies}. These techniques are valuable, but their gains are generally incremental and must be assessed together with their implementation cost.

The most consequential remaining PHY-layer opportunity is in MIMO transmission. The fundamental result in \cite{foschini} showed that the capacity of a point-to-point link can scale with the number of usable spatial degrees of freedom. In cellular networks, MIMO additionally provides array gain, interference suppression, and simultaneous transmission to multiple user equipments (UEs). Coordination between sectors can further mitigate inter-cell interference and improve spectral efficiency \cite{Valenzuela}. With the widespread use of active antenna arrays, equipping radio base stations (BSs) with tens or hundreds of antenna elements has become practical \cite{rel18mimo}. Although commercial deployments are not yet in the extremely large-array regime \cite{Rusek,massivemimo}, 6G is expected to continue this evolution to maintain and improve coverage and capacity, including in challenging mid-band deployments \cite{shafi2025industrialviewpointsrantechnologies}. Coordination between adjacent sectors has already shown promising gains and provides a practical stepping stone toward multi-transmission-reception point (TRP) operation \cite{huang}.

Critically, the importance of massive MIMO \cite{6736761} is not only the availability of more antenna elements, but also the scalability of channel state information (CSI) acquisition through time-division duplex (TDD) reciprocity. In TDD operation, a BS can estimate the high-dimensional channels of active UEs from uplink pilots and, after reciprocity calibration, reuse these estimates for downlink precoding and uplink combining. The pilot dimension therefore scales primarily with the number of simultaneously trained UEs and the channel coherence interval, rather than with the number of BS antenna elements. This principle is a central enabler of massive MIMO in TDD spectrum \cite{Rusek,massivemimo}. Reciprocity does not make CSI free: the propagation channel is reciprocal, whereas the end-to-end RF chains require calibration, and finite coherence time, pilot reuse, channel aging, and calibration errors remain practical limitations. A further challenge arises for cell-edge UEs. Their uplink pilot energy is constrained by UE transmit-power limits and large path loss, which can result in a low received pilot SNR at the BS. Consequently, the estimate of the reciprocal propagation channel can be noisy, particularly in the presence of pilot reuse and inter-cell interference, reducing the effectiveness of subsequent downlink precoding and uplink combining. Downlink CSI reference signal (CSI-RS) transmission and reporting in 5G new radio (NR) remain valuable for beam management, link adaptation, frequency-division duplex operation, and advanced multi-TRP transmission. However, CSI feedback should be viewed as complementary to, rather than a replacement for, reciprocity-based CSI acquisition in massive-MIMO TDD operation.

Distributed MIMO, including multi-TRP and cell-free MIMO architectures, extends these principles across geographically separated transmission points. When the participating BSs or TRPs operate with sufficiently accurate relative timing, frequency, and phase, they can act as a distributed array for coherent joint transmission, combining macro-diversity with coordinated interference management. Such phase-coherent operation is especially important for improving lower-tail service performance, where conventional small-cell deployments remain constrained by hard cell boundaries, shadowing, and inter-cell interference. The gains from network coordination are well established \cite{Valenzuela,huang}, but their practical realization depends on fronthaul or midhaul capacity and latency, transport jitter, timing/frequency/phase synchronization, distributed reciprocity calibration, CSI and user-data exchange, pilot coordination, and scalable scheduling. These requirements are not merely implementation details; they determine how much of the theoretical distributed MIMO gain can be achieved in a deployable network.

Alongside these spatial advances, artificial intelligence (AI) and machine learning (ML) create new opportunities for PHY-layer tasks such as channel estimation, equalization, CSI compression, and impairment compensation. AI-based neural receivers and CSI designs should be assessed against strong model-based baselines and according to their performance gain, overhead, computational complexity, robustness, and deployability. They can complement massive and distributed MIMO, but they are not the sole path beyond PHY-layer maturity.

This paper provides a realistic assessment of PHY-layer improvements that can enhance spectral efficiency and other performance measures, while accounting for complexity and practical deployment constraints. Section \ref{saturation} discusses measures of saturation and what is achievable in field deployments. Section \ref{gain} examines potential gains from constellation design, LDPC codes, reduced-complexity receiver architectures, waveform evolution, MIMO and CSI design, and neural receivers. Finally, Section \ref{sec:conclusion} concludes the paper.

\section{Metrics and Field Measurements}
\label{saturation}

\subsection{Spectral Efficiency Metrics}
\label{measures}

PHY layer saturation can be measured by the spectral efficiency and many of its flavors \cite{5741160, boccuzzi2025spectralefficiencyconsiderations6g}. There are multiple ways to define spectral efficiency arising from randomness in a wireless network, which, in turn, makes spectral efficiency a random variable \cite{6730658}. The spectral efficiency measures may include peak spectral efficiency, median spectral efficiency, cell-edge spectral efficiency, sum spectral efficiency, average area spectral efficiency, etc.

The mean signal-to-interference-plus-noise ratio (SINR) of a wireless cell may vary from -6 dB to 30 dB. The lower value corresponding to the cell edge and the higher value corresponding to the peak. Therefore, we have cell-edge spectral efficiency and peak spectral efficiency corresponding to these SINR values. MIMO systems result in an increase of the spectral efficiency.
For single-user MIMO (SU-MIMO) systems, at high SINR, the expected value of  spectral efficiency is directly proportional to the channel rank and is logarithmically dependent on the SINR. The channel rank varies over the cell and is in turn SINR dependent. For multi-user MIMO (MU-MIMO) systems, we would need to consider sum capacity, i.e., sum of the capacity realized by all the $K$ users. Simultaneous transmission to the $K$ users is achieved via precoding, such as zero forcing (ZF) precoding and block diagonalization \cite{Jindal_2005,1261332}. Therefore, a sum spectral efficiency based on the sum of the rates per user can be defined. 

The average area spectral efficiency  $C_{area}$ measured by  $bps/Hz/m^2$ is the average spectral efficiency over a wide area \cite{ITURM2410,775355}. Correspondingly, the area traffic capacity is the total traffic throughput served per geographic area (in $Mbps/m^2$). The throughput is the number of correctly received bits, i.e., the number of bits contained in the service data units delivered to layer 3 over a certain period of time. This concept was used in defining the minimum requirements of 5G but is also being considered for 6G \cite{ITURM2410}. Yet some other measures are energy efficiency measured in bits/Joule and its relationship to the sum throughput \cite{1091871}.

\subsection{Field Measurements}

The peak data rate is a key parameter in deciding the spectral efficiency of the PHY layer and deciding if it has reached saturation by comparing it with the Shannon limit.
When computing the peak data rate, it is assumed that the SNR is maximum and the channel offers full rank. For a single-user case, the maximum supported data rate in aggregated carriers in a band or band combination is computed following the calculation provided in Section 4.1.2 of \cite{3GPPTS38306}. Considering this, the maximum downlink data rate in case of single carrier without considering the network overhead can be computed as follows: 
\begin{equation}
    \label{predictedbestcaserate}
    R_{\textrm{max}} = q_{\textrm{max}}\hspace{1pt}
    r_{\textrm{max}}\hspace{1pt}
    \ell_{\textrm{max}}
    \left(N_{\textrm{RB}}
N_{\textrm{SC}}^{\textrm{RB}}\right)N_{\textrm{symb}}^{\textrm{slot}}\left(\frac{2^\mu}{T_{\textrm{sf}}}\right).
\end{equation}
Here $q_{\textrm{max}}$ is the maximum supported modulation order which takes the value of 8 or 10 in case of 256-QAM or 1024-QAM. Note that $r_{\textrm{max}}$ is the highest coding rate and is equal to $948/1024 = 0.93$ in 5G. Furthermore, $\ell_{\textrm{max}}$ denotes the maximum supported layers which is 4 in case of enhanced mobile broadband (eMBB) users and 8 for fixed wireless access (FWA); $N_{\textrm{RB}}$ denotes the maximum number of resource blocks (RBs) allocated in a given bandwidth $B$; $N_{\textrm{SC}}^{\textrm{RB}}$ is  the number of subcarriers per RB and is equal to $12$. Note that $N_{\textrm{symb}}^{\textrm{slot}}$ is the number of OFDM symbols per slot and is equal to $14$, $T_{\textrm{sf}}$ is the subframe duration and is equal to $1$ ms, and $\mu = 0, 1, 2, 3$ determines subcarrier spacing (SCS) value as $15 \times 2^\mu$ kHz. Parameters for Eq. (\ref{predictedbestcaserate}) are given in Table \ref{Tab:tbl}. It is worthwhile to see if the peak data rates as predicted by Eq. (\ref{predictedbestcaserate}) are achievable in practice.

\begin{table*}
%[!t]
    \centering
    \caption{Parameters to compute peak data rates in Eq. (\ref{predictedbestcaserate}).}
    \label{Tab:tbl}
    \begin{tabular}{|c|c|c|c|c|c|}
        \hline
            Frequency range & SCS (kHz) & Max BW (MHz) & $N_{\textrm{RB}}$ & $\ell_{\textrm{max}}$ & $R_{\textrm{max}}$ (Gbps)\\ \hline \hline
            410 - 7,125 MHz & 15 & 50 & 270 & 4 & 1.34 \\ 
            & 15  & 50  & 270 & 8 & 2.69 \\ 
            & 30  & 100 & 273 & 4 & 2.72  \\ 
            & 30  & 100 & 273 & 8 & 5.43  \\ \hline
            24.25 - 71 GHz & 60 & 200 & 264 & 4 & 5.26  \\ 
            & 60  & 200 & 264 & 8 & 10.51\\ \hline
            7.125 - 24.25 GHz & 30  & 400 & 1092 & 4 & 10.87  \\ 
            & 30  & 400 & 1092 & 8 & 21.74 \\
        \hline
    \end{tabular}
\end{table*}

In a practical system, the equation in (\ref{predictedbestcaserate}) needs to be scaled to account for guard bands, TDD factor, and control information overhead. In a $100$ MHz bandwidth, the guard band is 1.69 MHz, the TDD bandwidth reduction factor for a commonly used TDD pattern (DDDSU, with D being a downlink slot, S being a special slot, and U being an uplink slot) is on the order of $0.75$. The control information overhead for a 4-stream per-user transmission is approximately $3/14$ for $14$ OFDM symbols. However, it is difficult to precisely characterize a value for this as it is linked to implementation-specific characteristics of the system.

\begin{figure}[h]
		\centering
		\includegraphics[width=1.0\linewidth]{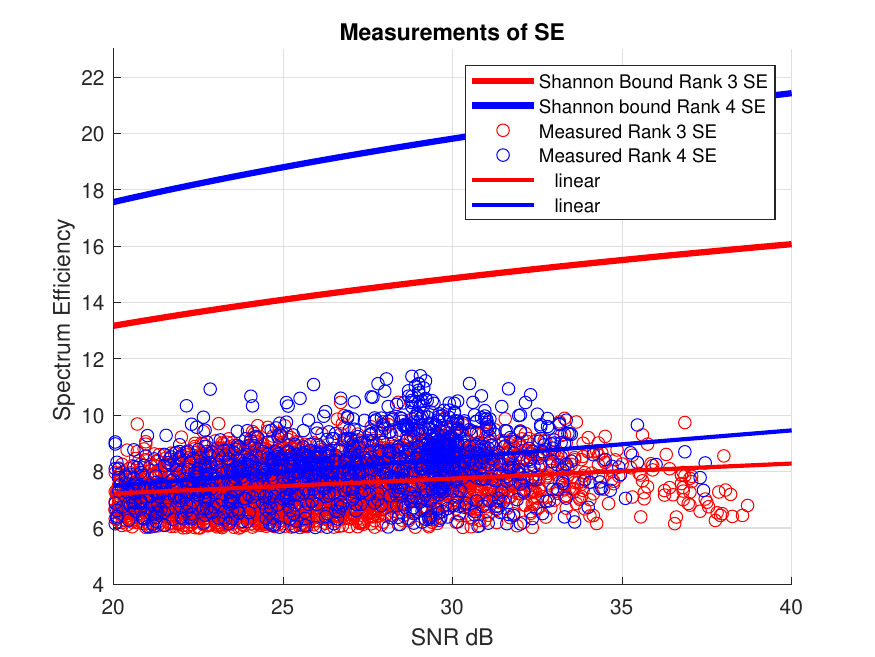}
		\caption{Measurements of spectral efficiency (SE) at high SNR via drive tests for a cluster of 5G cells using 100 MHz bandwidth and comparison with the capacity bound for Rank 3 and Rank 4.}
		\label{fig:drive_tests}
	\end{figure}

\begin{figure}[h]
		\centering
		\includegraphics[width=1.0\linewidth]{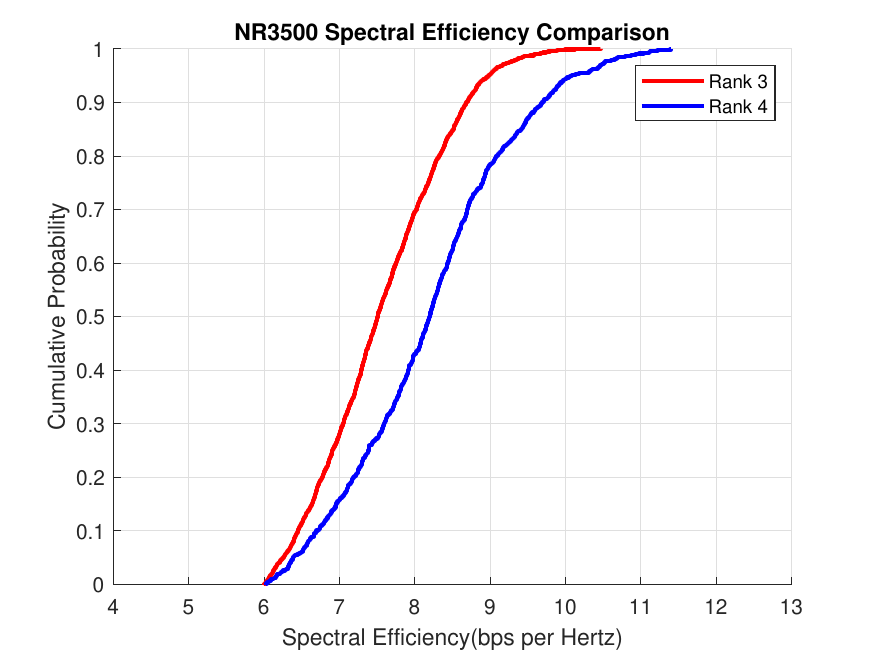}
		\caption{Measured CDFs of spectral efficiency at high SNR for Rank 3 and Rank 4 in a cluster of 5G cells using 100 MHz bandwidth.}
		\label{fig:CDF's}
	\end{figure}

Figure \ref{fig:drive_tests} shows real-world measurements of spectral efficiency for many clusters of 5G cells using 100 MHz bandwidth and 256-QAM for high SNR ranging from 20 dB to 40 dB. Both rank-3 and rank-4 data for SU-MIMO are shown and compared with the corresponding values of mean spectral efficiency from the capacity bound. \textbf{The gap between the capacity bound and measurements is due to TDD pattern, guard band, and control information overhead. If we account for these in Eq. (\ref{predictedbestcaserate}), the measurements show that the cells are operating close to the capacity bound.} The corresponding cumulative distribution functions (CDFs) of spectral efficiency are shown in Fig. \ref{fig:CDF's}. It is also noted that rank-3 occurrence is much more frequent than rank-4 case even though the SNR range is quite high.

\section{Potential Improvement Areas}
\label{gain}

\subsection{Signal Constellation Designs}
\label{modulation}

\begin{figure}
      \centering
      \includegraphics[width=1.0\linewidth]{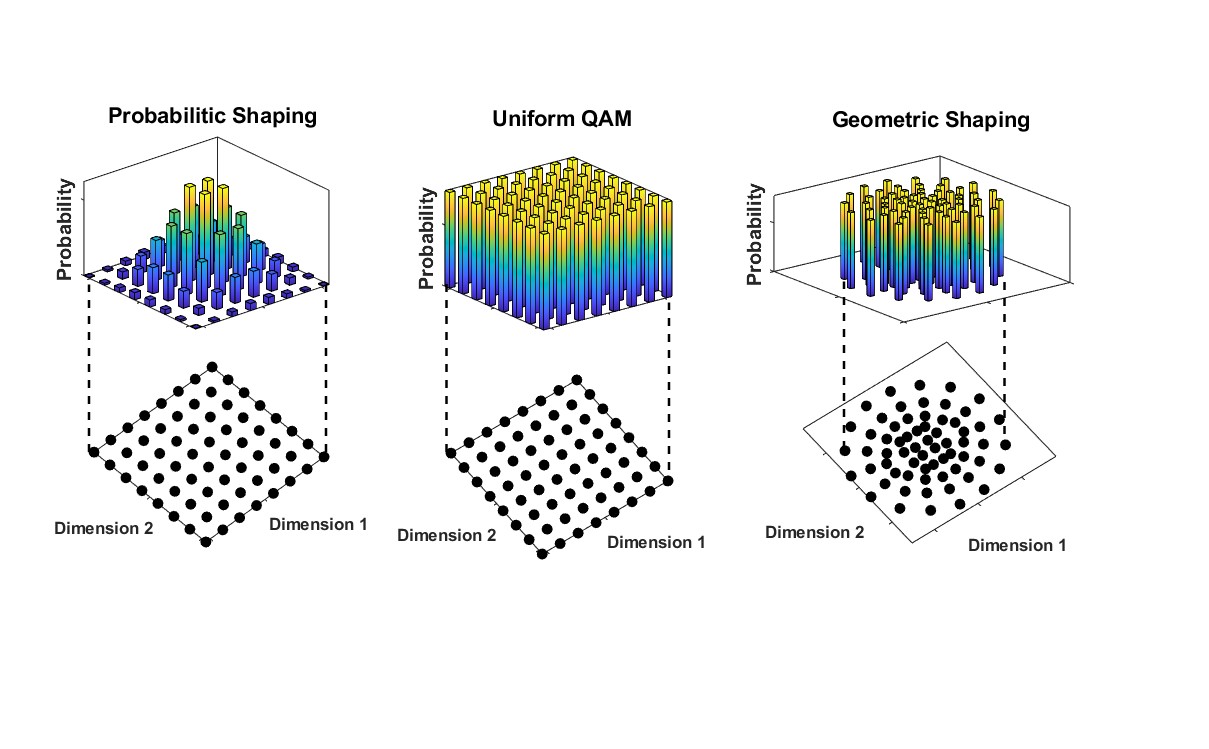}
      \caption{Illustration of uniform QAM, PS, and GS.}
      \label{fig:psgsshape}
  \end{figure}

In the conventional approach to the choice of a modulation scheme, each symbol is equally likely. Uniform QAM is widely used in 5G.  For a modulation of size $Q$, the supported bits per symbol $m$, called modulation order, is equal to $\log_{2}Q$. Peak modulation order is important for peak throughput. In 3GPP  Release 17, 1024-QAM is standardized for the downlink and 256-QAM for the uplink. Improving spectral efficiency can be achieved by even higher order QAM such as the use of 4096-QAM in the downlink. However, the very high SNR requirements for 4096-QAM and tighter RF linearity requirements make the use of 4096-QAM challenging in practice. The EVM requirements are dependent upon power amplifier nonlinearity, phase noise, and quantization loss in the transmitter. The latter two are also relevant for the receiver. The operating SNR for 1024-QAM is approximately $25$ dB and $36$ dB for 4096-QAM; clearly 36 dB SNR over a cell is not common. However, classical papers \cite{pasupathy,qureshi,Forney} show that non-uniform signaling can achieve an ultimate shaping gain of $\frac{\pi e}{6}$ or $1.53$ dB relative to uniform signaling but only when the modulation order is quite large. It is justified to ask: \textit{how to design signaling schemes that can achieve this ultimate shaping gain?} There are two approaches to shaping- GS and PS. The principle of uniform QAM, GS and PS is depicted in Fig.~\ref{fig:psgsshape}. 
  
In GS, non-uniform constellations employ optimized spatial distributions such as Gaussian distribution for the symbol points \cite{qureshi,1092061}. This adjusts the “shape” of the constellation to approximate the Gaussian signals and the Euclidean distance between adjacent points can be maximized. After bit-interleaved coded modulation (BICM) capacity was proposed, it is widely used for GS design and optimization \cite{Giuse1998BICM}. Several 8-ary and 16-ary constellations were evaluated using the capacity limit of BICM in \cite{Steph2003MOD}. The GS patterns from 16-QAM to 4096-QAM were designed for different application scenarios for broadcasting system in \cite{Jon2017MOD}. An efficient GS design was proposed with low complexity in which a subset of constellation points from a higher order QAM constellation is selected to form a GS constellation pattern with high BICM capacity in \cite{gcs}. Because of good performance, GS technology was adopted in the advanced television systems committee (ATSC) \cite{ATSC2017MOD,sonygcs}.

In PS, a constellation point $\textbf{r}$ is chosen with the probability $p\left(\textbf{r}\right) \sim exp \left( -\lambda\left(\lVert \mathbf{r} \right\rVert^2 \right)$, where $\lambda$ is an optimization parameter of the Maxwell--Boltzmann (M-B) distribution \cite{pasupathy,qcom}. It is noted that in PS the geometric structure remains unchanged. The parameter $\lambda$ can be adapted to different SNR and modulation orders. For example, with 256-QAM a $1$ dB shaping gain can be realized\cite{7307154} with an appropriate choice of $\lambda $. When $\lambda = 0$, a uniform distribution is obtained. A limiting constellation only selects the inner most points. As $\lambda \rightarrow \infty$, the bit rate as well as average energy are reduced, and points with large energy are selected less frequently. A value of $\lambda $ in between the extreme limits aims to achieve a balance between the inner most points and points with large energy. PS achieves the shaping gain by adjusting the probability of each constellation point to approximate the M-B distribution. This is realized  by statistically transmitting more low-power constellation points, thereby reducing the average transmission power of the signal. Under the same average power constraint, this approach in effect expands the distance between constellation points and thus leads to greater noise tolerance.

\begin{figure}
    \centering
    \includegraphics[width=1\linewidth]{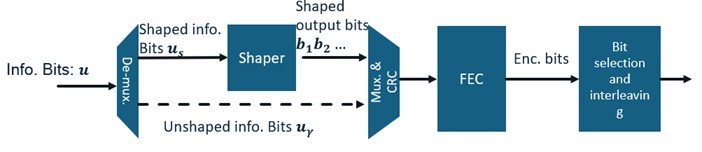}
    \caption{Transmission chain with PS.}
    \label{fig:txchainps}
\end{figure}

Concerning the implementation, there is no impact on the whole transmit chain other than modulation block itself for GS, since only the values of the constellation points are changed. Compared with GS, PS has large impact on transmit chain. An example of the code-block level processing flow for PS is depicted in Fig.~\ref{fig:txchainps}. Typically, PS is applied separately to the I and Q branches. Therefore, we can describe the procedure of Q-QAM modulation as a $\sqrt{Q}$-ary amplitude shift keying (ASK) signal, where $\sqrt{Q}$ denotes the number of constellation points associated with one I/Q branch of the QAM signal. For the case of 256-QAM, i.e., $Q = 256$, it is $16$-ary ASK signal for each branch. To reformulate into conventional QAM signal, we assume that the transmitted signals on I/Q branches are associated with the even and odd indexed ASK symbols, respectively.

The input information bit vector $\boldsymbol{u}$ is partitioned into two streams, where $\boldsymbol{u_s}$ is the shaped information bit vector and $\boldsymbol{u_\gamma}$ represents the unshaped information bit vector. The shaped branch $\boldsymbol{u_s}$ is processed by a shaper module, which outputs a shaped bit sequence $\boldsymbol{b}$. The unshaped stream $\boldsymbol{u_\gamma}$ enables rate adaptation, allowing the system to support varying target SNRs and spectral efficiencies. The shaper takes $\boldsymbol{u_s}$ as inputs and utilizes distribution matching to transform bits into probabilistically shaped bit sequence $\boldsymbol{b} = [\boldsymbol{b_1}, \boldsymbol{b_2}, \ldots]$. The unshaped bits $\boldsymbol{u_\gamma}$ and shaped bit sequence $\boldsymbol{b}$ are concatenated and cyclic redundancy check (CRC) bits are appended to the code block prior to forward error correction (FEC) encoding. 

\begin{figure}
    \centering
    \includegraphics[width=1\linewidth]{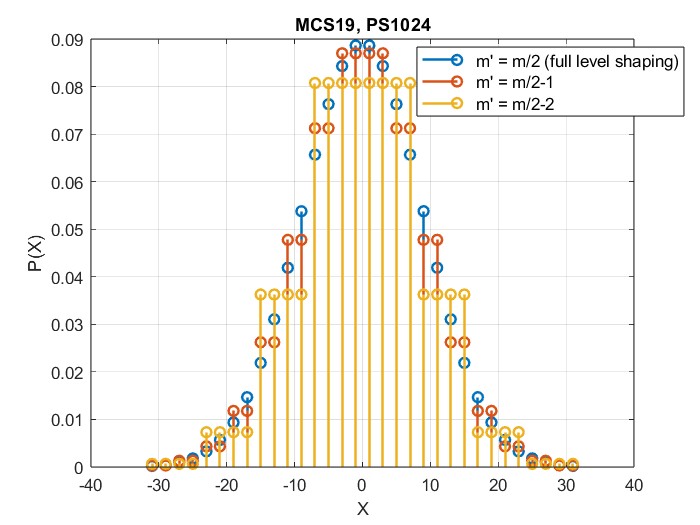}
    \caption{Transmission probability with partial shaping.}
    \label{fig:partitionprob}
\end{figure}

For a higher order QAM, $m/2$ bits are mapped to each I/Q symbol, where the most significant bit (MSB) is mapped to the sign of the modulation symbol, and the last $m/2 - 1$ bits determine the amplitude level of the modulation. The probabilistic shaping gain is dominated by the first subset of bits of the $m$ bits. Further complexity reduction can be achieved by configuring a proper parameter $m' < m/2 - 1$ to be probabilistically shaped for a given modulation order. In other words, only a subset of bit levels needs to be shaped. The remaining $m/2 - 1 - m'$ bit levels can be unshaped and uniformly distributed. In the demodulator, partial shaping also uses the configuration for full-level shaping, meaning it relies on the full-level M-B parameter $\lambda$ and the corresponding scaled power. In this case, the probability distribution could be staircase-like, as shown in the Fig.~\ref{fig:partitionprob}. It can be verified that the capacity can be close to the case with $m' = m/2-1$. We refer to this shaping scheme as \textit{low-dimensional probabilistic shaping}, and for higher order QAM (e.g., 256-QAM and 1024-QAM), we use low-dimensional probabilistic shaping for the performance and complexity evaluations, unless otherwise specified.

\begin{figure}
    \centering
    \includegraphics[width=1\linewidth]{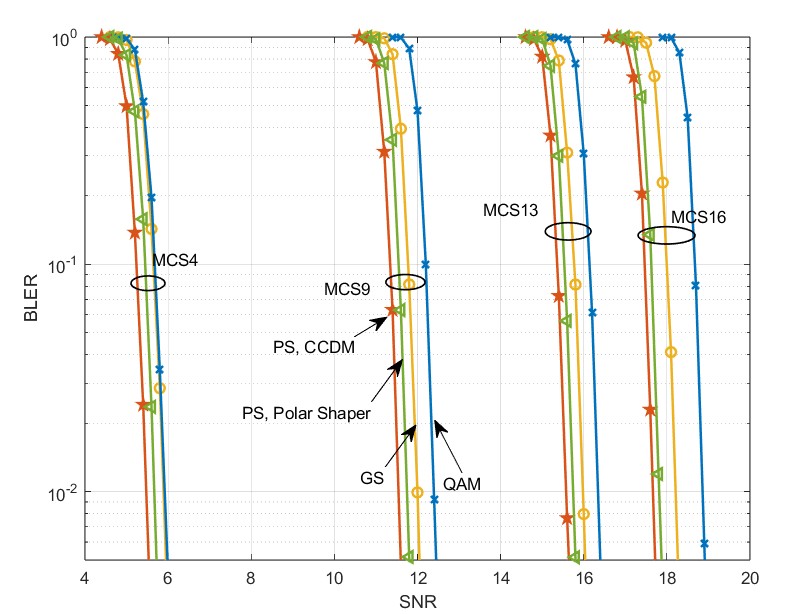}
    \caption{BLER performance of GS and PS for MCS levels of 4, 9, 13, 16.}
    \label{fig:blermcs4to16}
\end{figure}

\begin{figure}
    \centering
    \includegraphics[width=1\linewidth]{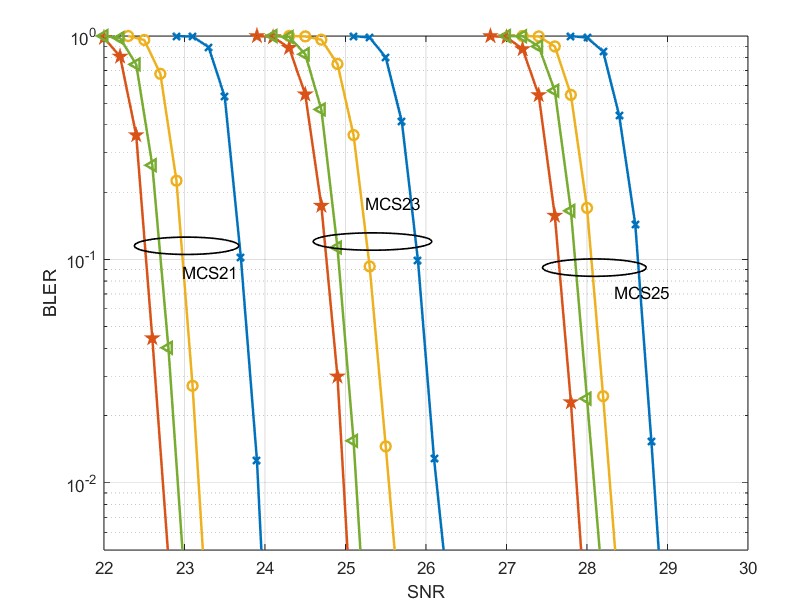}
    \caption{BLER performance of GS and PS for MCS levels of 21, 23, 25.}
    \label{fig:blermcs21to25}
\end{figure}

The block error rate (BLER) performance vs. SNR for PS and GS under various modulation and coding scheme (MCS) levels is shown in Fig.~\ref{fig:blermcs4to16} and Fig.~ \ref{fig:blermcs21to25}, respectively. The shaping gains of PS and GS over uniform QAM are listed in Table \ref{tab:shaping gain}. Considering Fig.~\ref{fig:blermcs4to16} and Fig.~\ref{fig:blermcs21to25}, it is seen that PS achieves the best performance in all configurations. The performance gaps relative to uniform QAM become large with modulation order increasing and are more than $1$ dB at modulation orders larger than 6. The gains of GS over uniform QAM are always less than PS for all the cases. 

\begin{table}[h]
\centering
\caption{Shaping gain of PS \& GS over uniform QAM.}
\label{tab:shaping gain}
\begin{tabular}{|l|c|c|}
\hline
\multirow{2}{*}{\textbf{Modulation}} & \multirow{2}{*}{\textbf{PS with maximum $m'=2$}} & \textbf{GS, 1D for 1K} \\ & & \textbf{2D for 16/64/256} \\ \hline
16QAM   & 0.4--0.5 dB & No gain     \\ \hline
64QAM   & 0.8--1 dB   & 0.4--0.5 dB \\ \hline
256QAM  & 1.0--1.4 dB & 0.5--0.7 dB \\ \hline
1024QAM & 1.0--1.4 dB & 0.5--0.7 dB \\ \hline
\end{tabular}
\end{table}

\subsection{Low-Density Parity-Check Codes}
\label{coding}

Because of excellent performance, LDPC codes have been widely adopted for many standards such as DVB, WiFi, WiMAX, and ATSC 3.0. During the standardization of 5G NR, significant amounts of effort were spent to study and compare the performance and efficiency among different family of channel codes, including (enhanced) turbo codes, LDPC codes, convolutional codes, and polar codes. At the end of the 5G study, LDPC codes and polar codes were selected as the corresponding channel codes to use for data channel and control channel, respectively. This decision was made based on the superior performance as well as energy/cost efficiency of these two families of codes, in the corresponding regimes of interest.

For 6G systems, the peak throughput target is expected to increase significantly beyond that of 5G. For eMBB use cases, a peak UE throughput of 36~Gbps may be required in the first 6G release \cite{3gpp38914}. Supporting such high throughput motivates the design of enhanced LDPC codes with improved performance at a small number of decoding iterations, thereby enabling higher area efficiency and decoder throughput than the LDPC codes used in 5G NR.

A key limitation of 5G NR LDPC codes in the small-iteration regime is the use of punctured systematic nodes. In particular, 5G NR LDPC codes employ two punctured nodes with higher degrees than the remaining columns. This increases the effective average degree of the code and improves performance when decoding is performed with a large number of iterations. However, in the small-iteration regime, multiple punctured nodes may slow convergence because the decoder must first recover the punctured bits. Reducing the number of punctured nodes can therefore improve performance at low iteration counts, although it may lead to performance loss in the asymptotic regime.

\begin{figure}
    \centering
    \includegraphics[width=1\linewidth]{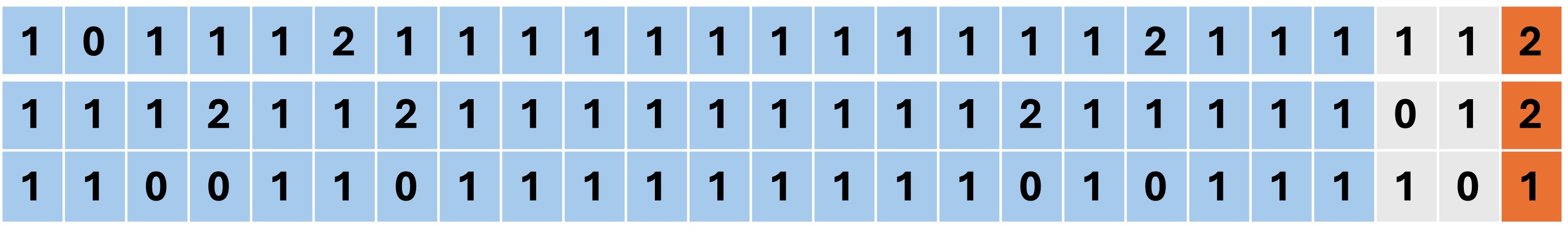}
    \caption{The proposed BG3 with core rate $R_{core}=22/24$.}
    \label{fig:bg3_core}
\end{figure}

Meeting both low-iteration and asymptotic performance goals is challenging when the base graph (BG) is restricted to single-edge connections, as in 5G NR BG1 and BG2. To address this issue, we may consider a new BG design (referred to as BG3) with a single punctured node and double edges. Using a single punctured column reduces the number of unknown bits and accelerates convergence, while double edges at the punctured node increase its total degree and improve asymptotic performance. The proposed design for BG3 for fully systematic LDPC is depicted in Fig.~\ref{fig:bg3_core} with a core size of \(3 \times 25\). The designed coding rate of the core $ R_{\mathrm{core}} = \frac{22}{24}$. We allow double edges not only in the punctured column but also in other locations. This provides degrees of freedom to achieve better trade-off between performance and decoding complexity. The number of information bits $K_b$ and the maximum lifting size $Z_{\max}$ are set as 22 and 384, respectively. These values are the same as 5G NR. The first 22 columns are systematic information bits while the last three columns are core parity columns. The 25-$th$ column is punctured when the coded block is transmitted. The target of this design is to optimize the performance at small and medium numbers of iterations, while maintaining very small loss relative to 5G BG1 at very large iterations.

\begin{figure}
    \centering
    \includegraphics[width=1\linewidth]{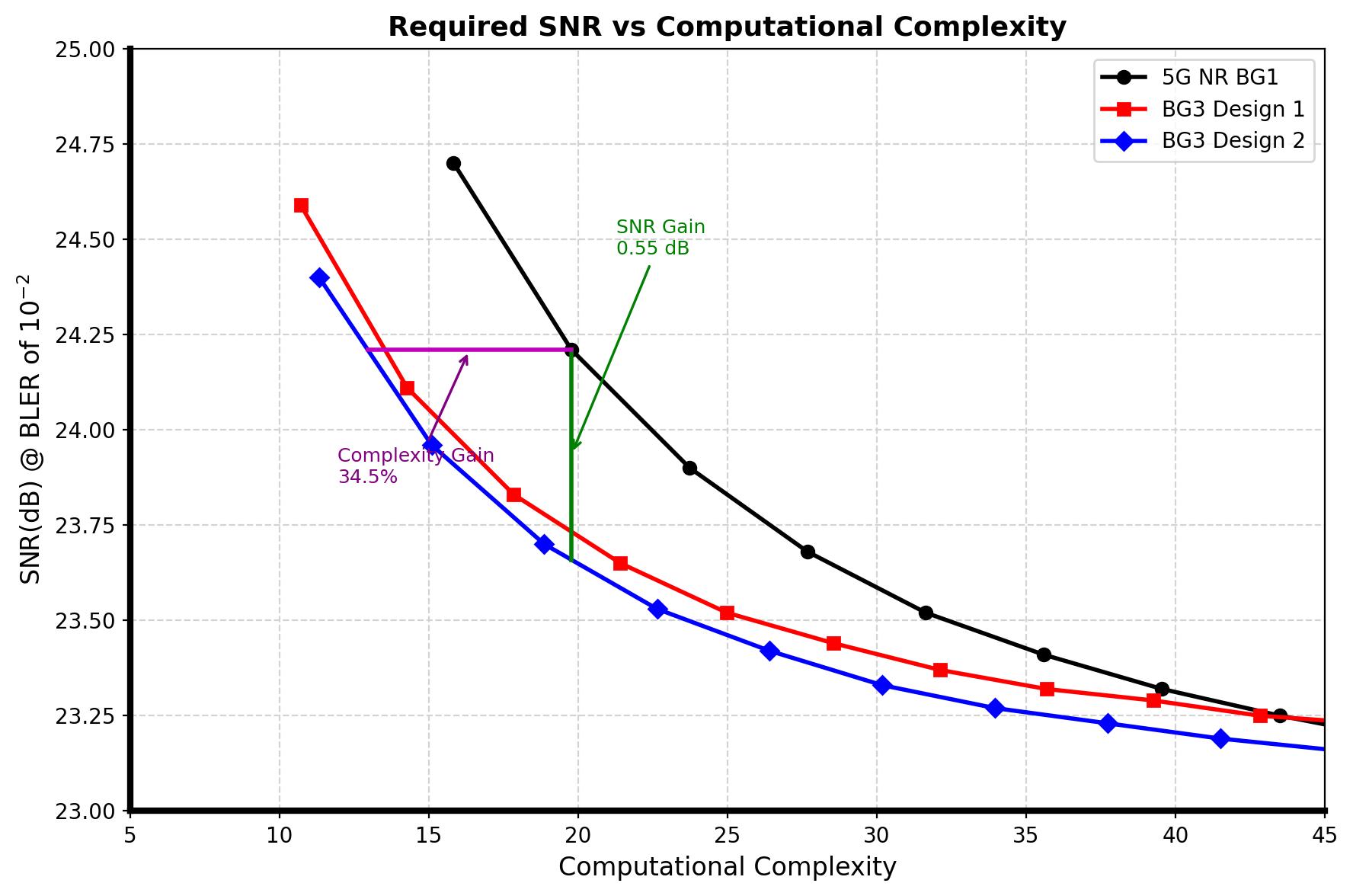}
    \caption{Performance comparison for Design 1, 2, and NR BG1 at BLER of $10^{-2}$.}
    \label{fig:PercomBLERm2}
\end{figure}

\begin{figure}
    \centering
    \includegraphics[width=1\linewidth]{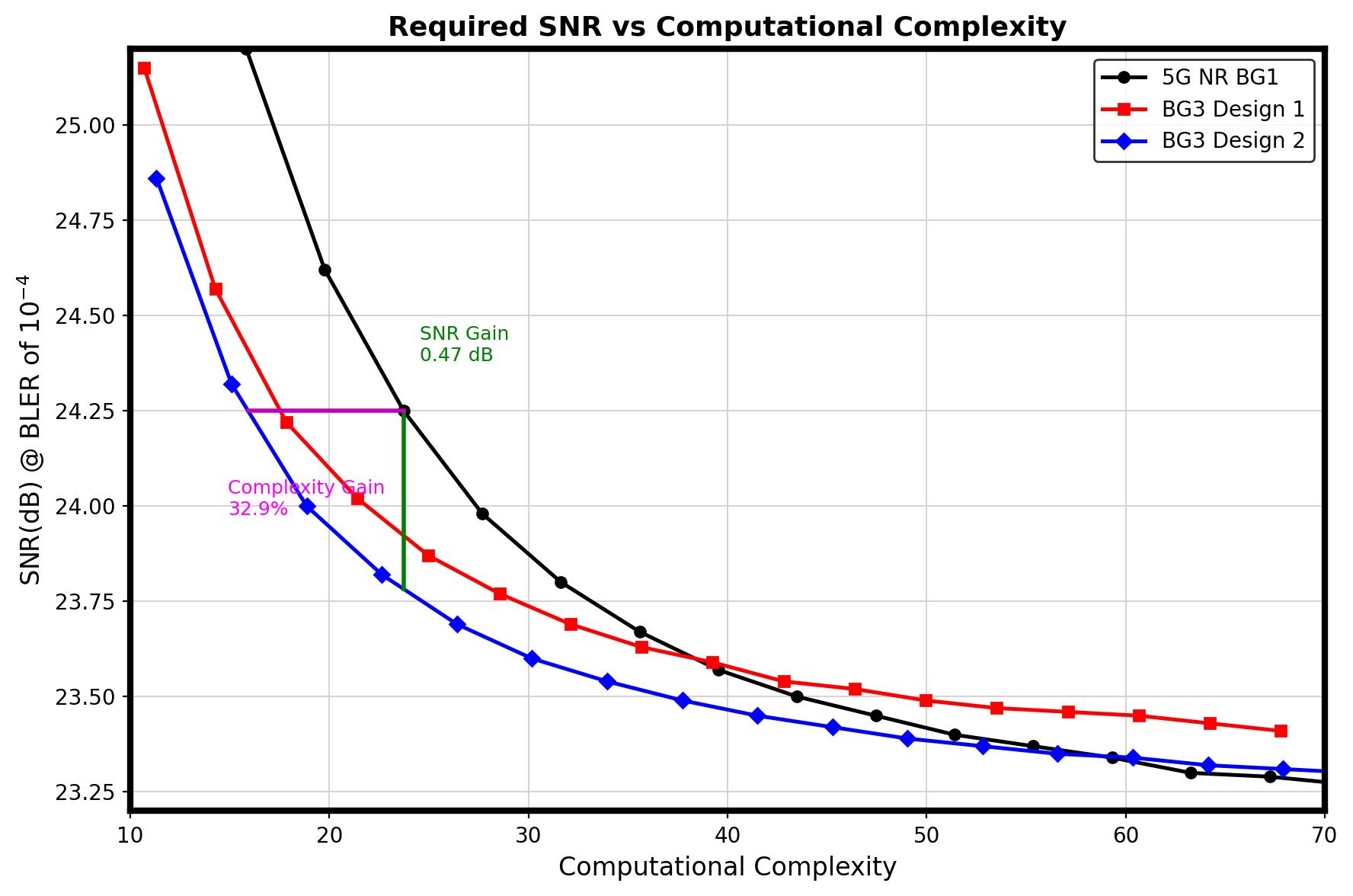}
    \caption{Performance comparison for Design 1, 2, and NR BG1 at BLER of $10^{-4}$.}
    \label{fig:PercomBLERm4}
\end{figure}

There are two designs of BG3, whose details can be found in ~\cite{Qualcoding2026}, ~\cite{Qualc2602}. The performance comparison between Design 1, Design 2, and NR BG1 is depicted in Fig.~\ref{fig:PercomBLERm2} under the additive white Gaussian noise (AWGN) channel with 256-QAM at the BLER target of $10^{-2}$. The simulation and algorithm details can be found in  ~\cite{Qualcoding2026}. The code block size is 8448 bits and code rate is $885/1024$. The horizontal axis, computational complexity, is defined as
\begin{equation}
    \text{Complexity} = \frac{N_{\text{iter}} \times N_{\text{ones}}}{N_{\text{CB}}},
\end{equation}
where $N_{\text{iter}}$ is the number of iterations required to achieve the
target BLER, $N_{\text{ones}}$ is the number of ones in the lifted parity-check
matrix, and $N_{\text{CB}}$ is the code block size. It is seen that the required SNR decreases with the increase of computational complexity for all the three BGs. The performance of Design 1 and Design 2 is better than NR BG1 at low complexity and comparable to NR BG1 at high complexity with large number of iterations. In particular, for the same complexity around 20, the SNR gain of Design 2 over NR BG1 is $0.55$ dB to obtain a BLER of $10^{-2}$. For the same SNR around $24.20$ dB, $34.5\%$ of the computational complexity is reduced for Design 2 compared to NR BG1. This implies that we can achieve either SNR gain or computational reduction gain with the new BG3 design at small number of iterations. The corresponding performance comparison for the target BLER of $10^{-4}$ is depicted in Fig.~\ref{fig:PercomBLERm4} with the same setup. The SNR gain and the computational reduction gain are $0.47$ dB and $32.9\%$, respectively. In addition, it can be seen that Design 1 outperforms corresponding NR code with slightly smaller gain than Design 2. In summary, both Design 1 and
Design 2 provide a better trade-off between performance and decoding complexity than NR~BG1. A
single punctured node with double edges helps achieve decent performance in the
medium/large iteration regime while providing significant gain in the small
iteration regime.

\subsection{Reduced-Complexity Maximum Likelihood Receiver Architectures}
\label{receiverarchitectures}

As the number of MIMO layers and transmit chains increase, decoding complexity in a UE increases very substantially. 
The quality of soft outputs produced by the demapper is a foundational
determinant of the overall link performance. Unlike hard-decision outputs,
soft-output demappers furnish the channel decoder with log-likelihood ratios (LLRs) that quantify the reliability of each coded bit, enabling the following decoder like LDPC decoder to exploit the full iterative coding gain. In high-order MIMO
systems, however, exact maximum-likelihood (ML) LLR computation requires an
exhaustive search over an exponentially large symbol constellation space,
rendering it computationally intractable for practical implementations.
Sphere-decoding (SD) algorithms offer a near-ML soft-output solution
by restricting the search to a hypersphere of controlled radius, thereby
achieving ML-comparable LLR quality at a substantially reduced computational
cost \cite{Li_2011,4444760, 1408197,1459002}. When a PS constellation is applied to the transmitter, the uniform prior assumed by conventional SD is no longer valid;
the demapper must incorporate M-B distribution into its metric to preserve LLR accuracy. 

In this section, we consider a PS-aware SD with complexity reduction for MIMO systems; the detailed scheme is described in \cite{11154592}. The performance gain and the corresponding complexity reduction are summarized in  Table~\ref{tab:complexity} across two representative channel correlation regimes at a target BLER of $10^{-2}$. The antenna correlation can be generated according to Section 7.7.5.2 in \cite{ts901}. Specifically, the values of ($\alpha, \beta$) are set as (0, 0) and (0, 0.3874) for low correlation and medium correlation, respectively. It is shown that PS-aware SD outperforms conventional SD with more than $2$ dB for low antenna correlation and $3.5$ dB for medium antenna correlation. This is reasonable because the inter-layer
interference is more detrimental to the performance of MIMO
systems in case of medium antenna correlation. Therefore, the interference shaping gain from PS becomes larger as the antenna correlation increases.    

\begin{table}[!t]
  \centering
  \caption{Performance comparison between PS-aware SD vs. conventional SD.}
  \label{tab:complexity}
  \renewcommand{\arraystretch}{1.25}
  \scalebox{0.9}{
  \begin{tabular}{|l|c|c|c|}
    \hline
    \textbf{Scenario} & \textbf{SNR Gain (dB)} & \textbf{Node Reduction} & \textbf{LLR Quality} \\
    \hline
    Low correlation   & $>2.0$ & ${\sim}10\times$  & Near-ML \\
    Medium correlation & $>3.5$ & ${\sim}3\times$   & Near-ML \\
    \hline
  \end{tabular}}
\end{table}

The node reduction is defined as the ratio of average visited nodes between
the conventional SD and the PS-aware SD under identical sphere radii
initialization. The complexity reduction arises from two complementary mechanisms. The first one is prior-weighted pruning. Because M-B-shaped constellations assign exponentially lower probability to high-energy symbols, the $\lambda|s_i|^2$ penalty causes high-energy branches to accumulate large partial distances early in the tree traversal, triggering
radius-based pruning at upper layers before the expensive lower layers are
evaluated. This concentrates the tree search on low-energy, high-probability
subtrees and reduces the average number of visited nodes substantially. The second is interference shaping. For spatially correlated MIMO channels,
the off-diagonal elements of $\mathbf{R}$ couple the residuals across layers.
The M-B prior implicitly shapes the effective interference profile seen at each
layer, because the expected interference from a shaped layer $j>i$ is lower
than under a uniform prior. This reduces the residual variance $\mathbb{E}[|e_i|^2]$
and consequently tightens the effective sphere radius, further restricting the
search space. The combined effect is most pronounced in low-correlation
channels, where the prior-weighted pruning dominates, and remains significant
in medium-correlation scenarios where the interference-shaping contribution
compensates for the reduced pruning depth.

From an implementation perspective, the PS-aware sphere decoder requires only
one additional multiply-accumulate operation per tree node relative to
a conventional SD. The computation of $\lambda|s_i|^2$, which is a real-valued
scalar look-up, is easily absorbed into the branch-metric pipeline. No structural changes to the tree traversal, radius management, or LLR back-computation logic are required. This makes the scheme fully backward-compatible with existing SD hardware accelerators. The large gains position the PS-aware SD as a key enabler for 6G. First, the more than $2$--$3.5$ dB performance gain over conventional SD substantially closes the gap to the $1.53$ dB AWGN shaping gain bound. This demonstrates that the true benefit of PS in MIMO extends well beyond the classical scalar-channel result and is achievable with practical receivers. Second, the low complexity of the $3$--$10\times$ node reduction keeps the SD within the power envelope of advanced mobile system on chip (SoC). Last, the multiply-accumulate metric augmentation can be implemented on 5G NR decoder infrastructure for backward compatibility. This reduces the standardization risk and enables time-to-market advantages for ecosystems ahead of adoption
cycles that typically lag by one to two product generations.

\subsection{Waveforms}
\label{waveforms}

OFDM-based waveforms have been specified in 5G for both frequency range (FR) 1 and FR2 (up to 71 GHz). NR supports cyclic prefix (CP) OFDM in downlink and both CP-OFDM and discrete Fourier transform (DFT) spread OFDM (DFT-s-OFDM) in the uplink. OFDM-based waveforms are particularly well suited for broadband cellular systems due to their high flexibility in multiple access, excellent compatibility with MIMO and spatial multiplexing, and efficient fast Fourier transform (FFT) based implementations at both the transmitter and receiver. These characteristics allow OFDM-based systems to scale efficiently with bandwidth, antenna count, and numerology, while maintaining manageable computational complexity.

\begin{figure}[h]
		\centering
		\includegraphics[width=1.0\linewidth]{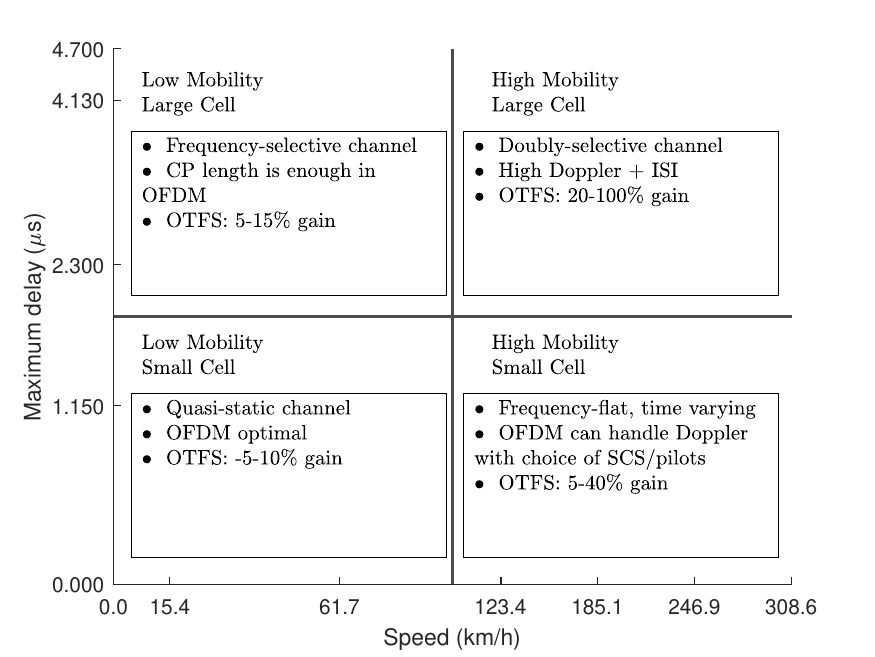}
		\caption{Spectrum efficiency comparison of OTFS and OFDM based on \cite{ref2}.}
		\label{fig:Waveforms}
	\end{figure}

It has also been agreed that for 6G, CP-OFDM will be the baseline waveform, and additionally DFT-s-OFDM will be supported in uplink~\cite{ref1}. The most important reasons for reusing the same waveforms are performance and implementation complexity. For 6G, alternative waveforms, including orthogonal time frequency space (OTFS) modulation and its Zak transform implementation, have attracted considerable attention in the literature~\cite{ref2}. OTFS is often highlighted for its potential advantages in highly time-varying channels, particularly those characterized by simultaneously large Doppler spreads and delay spreads. Several studies have evaluated the performance of OTFS against CP-OFDM under doubly selective channel conditions. However, from a practical perspective, the key is to consider the dominant cellular scenarios. Fig. \ref{fig:Waveforms} shows different communication scenarios and some estimate of potential performance gains based on~\cite{ref2}. It is important to note that the results in~\cite{ref2} are based on the extended vehicular A (EVA) channel model, which is not a representative channel model, as it emphasizes longer delay spread tails and does not consider any reference signal enhancements for CP-OFDM. Most practical communication scenarios are dominated by small cells and low mobility with typical maximum delays ranging up to \(100\text{--}300\,\mathrm{ns}\)~\cite{ts901}. In such cases, OTFS and CP-OFDM exhibit similar spectral efficiency, as also highlighted in~\cite{ref2}. Among the considered scenarios, OTFS could mainly provide more significant performance gains in large cells with high mobility, whereas commercial deployments are clearly dominated by small-cell and low-mobility environments.

Another important aspect is implementation complexity. Compared to CP-OFDM, the receiver complexity of OTFS remains fundamentally higher due to the loss of the diagonal channel structure. In OFDM, equalization is performed per subcarrier using a single complex multiplication, resulting in linear complexity
$\mathcal{O}(MN)$ for a frame comprising $M$ subcarriers and $N$ symbols. In contrast, conventional OTFS detection requires inversion of a dense \(MN \times MN\) channel matrix, leading to cubic complexity
$\mathcal{O}\big((MN)^3\big)$,
which is impractical.
Even with recent low-complexity algorithms exploiting sparsity or banded structures, such as conjugate gradient or message passing detectors, the complexity remains iterative, typically
$\mathcal{O}(k\,b\,MN)
\quad \text{or} \quad
\mathcal{O}(k\,MN \log MN)$,
where \(k\) denotes the number of iterations and \(b\) the effective channel spread~\cite{ref6}.
However, even with smaller values of \((k,b)\) as reported in~\cite{ref6}, the computational load remains multiple orders of magnitude higher compared to OFDM. Therefore, although low-complexity OTFS receivers significantly reduce complexity, they do not eliminate the inherent need for iterative multi-dimensional equalisation, which translates into increased processing latency and power consumption in practical implementations.

Specific waveforms such as OTFS have also been discussed for integrated communication and sensing (ISAC). However, ISAC implies that the underlying communication waveform should also be used for sensing; otherwise, the level of integration remains low and inefficient from a spectral efficiency perspective.
Furthermore, low integration levels may limit or even prevent simultaneous communication and sensing. Introducing a dedicated waveform for ISAC would require new transmit and receive processing chains, new multiplexing designs for coexistence on the same carrier, and increased standardization complexity to define new generation, mapping, reception, and measurement procedures~\cite{ref7}.

In the uplink direction, 6G will continue to support CP-OFDM and DFT-s-OFDM, as in 5G. DFT-s-OFDM is required for coverage enhancement due to its lower peak-to-average power ratio (PAPR). 5G NR already supports frequency-domain spectrum shaping (FDSS) to reduce PAPR for \(\pi/2\)-binary phase shift keying (BPSK) and quadrature phase shift keying (QPSK). FDSS operates on the DFT-s-OFDM waveform by applying a shaping filter in the frequency domain, enabling effective PAPR reduction with low implementation complexity.
Additional PAPR reduction techniques are being studied for 6G, particularly as 6G moves towards higher frequencies (e.g., around \(7\,\mathrm{GHz}\)). Promising techniques under discussion include FDSS with or without spectrum extension for QPSK, FDSS with spectrum truncation for \(\pi/2\)-BPSK, and crest factor reduction for QPSK~\cite{ref8}.
These techniques primarily target the spectral efficiency region between \(0.25\) and \(1.3\,\mathrm{bit/s/Hz}\), and may offer up to approximately \(1.5\,\mathrm{dB}\) coverage gain over 5G NR low-PAPR waveforms.

\subsection{MIMO}
\label{MIMO}

MIMO has been a key source of spectral efficiency improvements since 4G. SU-MIMO capability is now available in all commercial systems. If a BS
has $M$ antennas and a UE has $N$ antennas where $N\leq{}M$, the BS can transmit a total of up to $N$ streams to the UE. The $N\times{}M$ SU-MIMO propagation channel is denoted by $\textbf{H}\in{}\mathcal{C}^{N\times{}M}$. The system is assumed to operate over a long-term SNR denoted by $\rho$, which naturally is a function of the path loss, shadow fading, etc. These assumptions lead to a SU-MIMO spectral efficiency \cite{foschini}:
\begin{equation}
    \label{SUMIMO}
    R_{\textrm{SU}} = \log_{2}\left(\textrm{det}
    \left|\mathbf{I}_{N\times{}N}+\rho
    \mathbf{HH}^{H}
    \right|\right).
\end{equation}
The approximate value of Eq. \eqref{SUMIMO} at high SNR (high values of $\rho$) is given by $r\log_2(\rho)$, where $r$ is the rank of the SU-MIMO channel, also known as the degrees of freedom of the channel upper bounded by the $\min(N,M)$ \cite{TSE}. On the other hand, at low SNR (low values of $\rho$), Eq. \eqref{SUMIMO} can be approximated as $\rho{}\log_{2}(e) \textrm{tr}(\mathbf{HH}^H)$ \cite{TSE}.

Building on this, the case of MU-MIMO can now be analyzed. Assume that we have $K$ users in total, each capable of receiving $N$ streams, being served by a BS with $M$ transmit antennas, such that $M\geq{}KN$. In the classical MU-MIMO case, it is assumed that $N=1$. We assume that the $K$ UEs are served within the same time-frequency resource and hence the spatial domain is used via spatial and linear precoding to ensure that inter-user interference is reduced or nulled. In this case, the sum spectral efficiency across the $K$ UEs is up to $K$ times the corresponding single-user spectral efficiency \cite{Rusek}. Further increases in spectral efficiency  warrant the consideration of distributed MIMO (D-MIMO) BSs \cite{GLee}. Here, a BS is distributed into multiple $G$ BSs within the same cell, and all serving the same UEs. The multiple BSs could be on the same tower (co-located) or distributed over the cell. The former case is easier to implement than the latter \cite{huang}. Recently, D-MIMO (i.e., multi-TRP) has been considered as MIMO evolution in 5G-Advanced and now also in 6G. Assuming there are $G$ distributed BSs co-located or distributed over a cell, then the spectral efficiency will increase up to  further $G$ times.

D-MIMO serves as a representative model for cell free MIMO \cite{cellfree}.
The spectral efficiency gains can be summarized as shown in Table \ref{Tab:test2} \cite{Jindal_2005}.
\begin{table}

    \centering
    \caption{Spectral efficiency gains at asymptotic high SNR.}
    \label{Tab:test2}
    \begin{tabular}{|c|c|c|}
        \hline
             \textbf{SISO or MIMO Variant}   &  \textbf{Spectral Efficiency Gain}\\
        \hline \hline
            SISO 
            &  Baseline \\
        \hline
            SU-MIMO 
           & $\min(M,N)$ \\
        \hline
            MU-MIMO & 
            $KN$ where $KN\leq{}M$ \\ \hline
            D-MIMO & $GKN$, where $GKN\leq{} GM$\\
        \hline
    \end{tabular}
\end{table}
Whilst D-MIMO shows that the gain over MU-MIMO increases $G$ times, there are practical considerations that might limit this gain. For example, if the D-MIMO BSs are co-located, then only those UEs that have overlapping coverage from the distributed BSs (such as adjacent sectors, in which case $G$ is at most 2) will experience a D-MIMO gain. If the distributed MIMO BSs are distributed over a cell, the cost of fiber optic links to carry fronthaul traffic will place a limit. Also, it is noted that the gain of D-MIMO relative to centralized MIMO should be using $GM$ antennas in the centralized MIMO case, but this will increase the size of the central BS antenna and hence an unacceptable form factor. 

\begin{figure}
    \centering
    \includegraphics[width=1\linewidth]{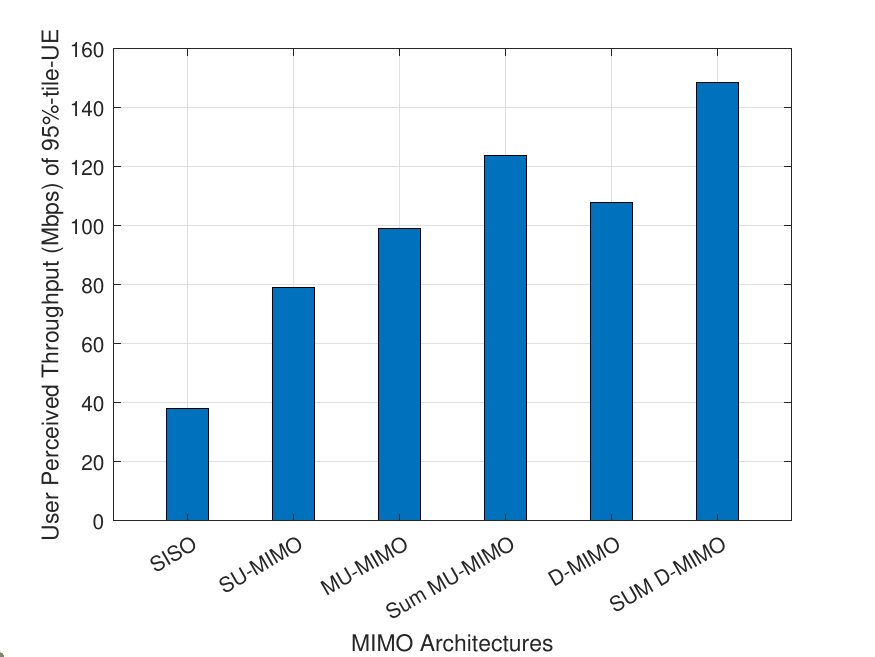}
    \caption{User perceived throughput (UPT) of 95\%-tile UE under different MIMO architectures: SISO, SU-MIMO, MU-MIMO, and D-MIMO.
 }
    \label{fig:MIMO simulations}
\end{figure}

While Table \ref{Tab:test2} summarizes the theoretical spectral efficiency at asymptotically high SNR for various MIMO architectures, realizing these ideal gains in a deployed network is highly dependent on practical considerations. Factors such as inter-cell interference, imperfect CSI, and the overhead required for multi-site coordination in addition to maximum code rate, link adaptation, and hybrid automatic repeat request (HARQ) process constrain the achievable capacity. To capture these realistic network dynamics, Fig. \ref{fig:MIMO simulations} shows system level simulation (SLS) results indicating the user perceived throughput (UPT) for the 95th-percentile UE across four deployment scenarios: SISO, SU-MIMO, MU-MIMO, and D-MIMO, where the detailed SLS assumptions follow 3GPP evaluation methodologies described in \cite{ts901,ref10}. The scenario is urban macro. The numbers of antenna ports at BS and UE are 64 and 4, respectively. Other key simulation assumptions include a hexagonal cell layout with 57 cells and 30 UEs per cell, 4 GHz carrier, 30 kHz SCS, 20 MHz bandwidth, CSI periodicity of 20 slots, scheduling delay of 4 slots, file transfer protocol (FTP) traffic model 3 with 30\% resource utilization. For the SISO case, 1 BS antenna port and 1 UE antenna port are used; for the SU-MIMO case, up to 4 layers are transmitted; for MU-MIMO case, up to 12 layers in total for scheduled UEs are used; for D-MIMO intra-site inter-cell scenario, a multi-TRP set consists of $G=3$ TRPs.

By focusing on the 95th-percentile UE, the evaluation shows the peak spatial multiplexing performance experienced by users under favorable channel conditions. The baseline SISO scenario demonstrates the fundamental throughput limit without spatial degrees of freedom. The transition to SU-MIMO yields a substantial throughput gain, directly reflecting the spectral efficiency gains of utilizing multiple spatial streams for a single user. Furthermore, MU-MIMO extends this gain by effectively spreading multiple users in the spatial domain within the same time-frequency resources. Finally, D-MIMO achieves the highest peak throughput. By leveraging macro-diversity and coordinated joint transmission from geographically distributed BSs, D-MIMO mitigates inter-sector interference and provides the maximum spatial degrees of freedom and throughput. Note that UPT is defined as the total size of packets successfully transmitted over the time taken to fully transmit the packets (i.e., active time), which 
 can be calculated from the time instance that a packet is in buffer to the time the packet is successfully and completely transmitted. Depending on the MIMO scheme and scheduler, the active time can be reduced or increased. For example, for the case of MU-MIMO as opposed to SU-MIMO, it can reduce the active time for more UEs in the cell by additionally transmitting packets to multiple UEs, which results in a UPT gain in the end, as seen in the results. On average, the number of simultaneous active UEs in the MU-MIMO simulation is 1.25 and in the case of D-MIMO it is 1.375. Therefore, sum capacity for these two cases is appropriately adjusted. 

Although the empirical UPT gains do not scale perfectly linearly with the number of antennas or TRPs (due to the aforementioned SLS constraints), the clear stepwise throughput gains in Fig. \ref{fig:MIMO simulations} validate the performance potential of advanced MIMO topologies. Consequently, as the network evolves toward D-MIMO and multi-TRP configurations, managing the exponentially growing CSI feedback overhead becomes the next critical challenge. This necessitates the development of highly optimized, unified quantization schemes for 6G, which are evaluated in the subsequent subsections.

\subsubsection{MIMO Codebooks}
As the number of antenna ports increases  to achieve the theoretical capacity gains of massive MIMO in 6G systems, the overhead associated with CSI feedback becomes a critical bottleneck \cite{shafi2025industrialviewpointsrantechnologies}. Hence, efficient CSI quantization and feedback mechanisms are needed to maintain the spectral efficiency gains without consuming excessive uplink resources conveying the CSI.

Here we discuss a new paradigm, termed the unified fixed codebook (UFC), to address the CSI feedback challenges. Unlike the 5G Release-16/19 eType-II codebook \cite{shafi2025industrialviewpointsrantechnologies}, which relies on layer-common spatial-domain basis vector selection (typically restricting the number of spatial-domain basis vectors to $L \in \{{2,4,6}\}$), the proposed UFC introduces layer-specific spatial-domain basis vector selection. Crucially, it allows for smaller values such as $L=1$ (or 2). By doing so, $L=1$ effectively subsumes the role of a low-resolution codebook (comparable to 5G Type-I codebook), while $L > 1$ operates as a high-resolution codebook (comparable to eType-II codebook), thereby unifying the two feedback mechanisms into a single design framework. In addition, the UFC incorporates frequency-domain (FD) compression (analogous to the eType-II codebook) to maximize compression gain across all operating scenarios, regardless of whether a low- or high-resolution configuration is applied. Furthermore, the UFC reduces feedback overhead by eliminating the bitmap required for selecting non-zero coefficients (NZCs), thereby avoiding unpredictable loss in reconstruction accuracy.

\begin{figure}
    \centering
    \includegraphics[width=1\linewidth]{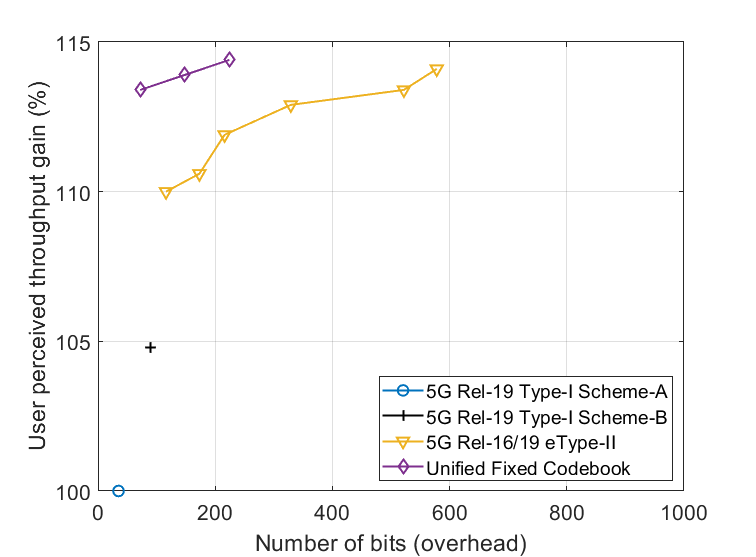}
    \caption{UPT vs. overhead trade-off for 5G NR CSI codebooks and unified fixed codebook in single-TRP scenario.}
    \label{fig:UPT vs overhead}
\end{figure}

\begin{figure}
    \centering
    \includegraphics[width=1\linewidth]{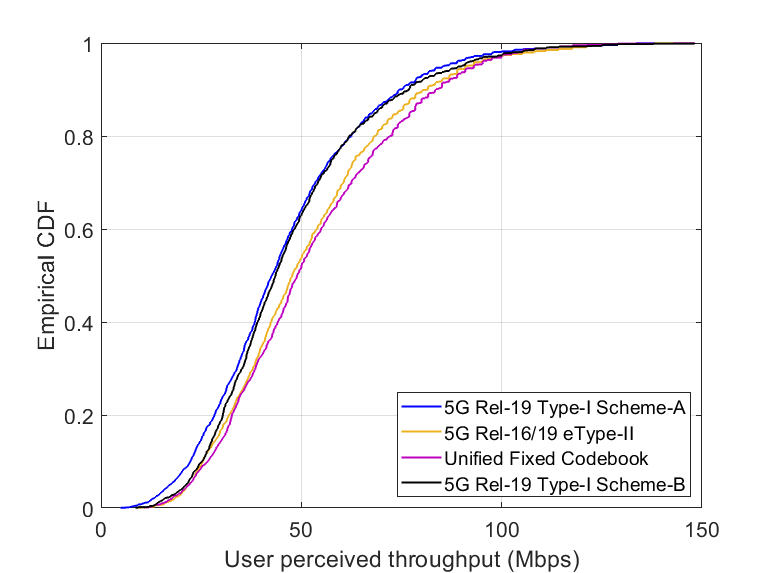}
    \caption{CDF of UPT for 5G NR CSI codebooks and unified fixed codebook in single-TRP scenario.}
    \label{fig:UPT Cdf}
\end{figure}

To demonstrate the potential of this paradigm, Figs. \ref{fig:UPT vs overhead} and \ref{fig:UPT Cdf} depict the performance of the proposed 6G UFC against the existing 5G CSI codebook schemes, including 5G Release-19 Type-I (Scheme-A and Scheme-B) and 5G Release-16/19 eType-II, via SLS evaluations. Assumptions for the SLS evaluations follow 3GPP evaluation methodologies described in \cite{ts901,ref10}. Fig. \ref{fig:UPT vs overhead} illustrates the trade-off between CSI feedback overhead (measured in bits) and UPT gain (\%). It is observed that the UFC achieves a substantial throughput gain ranging from 113\% to 114\% while requiring less than 200 bits of overhead. In contrast, the 5G Rel-16/19 eType-II scheme requires significantly larger payload, scaling up to nearly 600 bits to attain comparable performance. Fig. \ref{fig:UPT Cdf} presents simulated CDFs of the UPT (Mbps) for all users in the simulation layout. It is observed that at any UPT percentile, the UFC attains higher UPT compared to the legacy 5G Type-I and eType-II codebooks. This confirms that the UFC not only drastically reduces the uplink feedback overhead but also enhances UPT performance across a wide range of user channel conditions.

\subsubsection{Multi-TRP}

\begin{figure}
    \centering
    \includegraphics[width=1\linewidth]{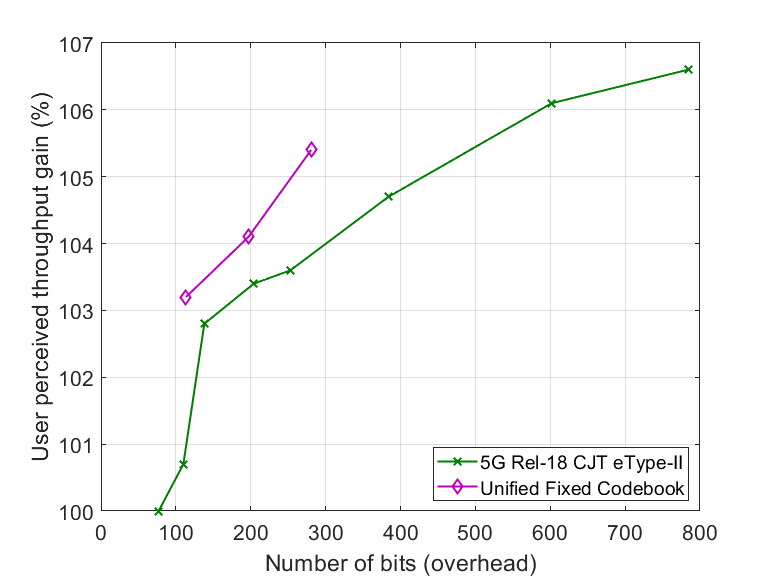}
    \caption{UPT vs. overhead trade-off for 5G NR CSI codebook and unified fixed codebook in multi-TRP CJT scenario.}
    \label{fig:UPT multi-TRP}
\end{figure}

\begin{figure}
    \centering
    \includegraphics[width=1\linewidth]{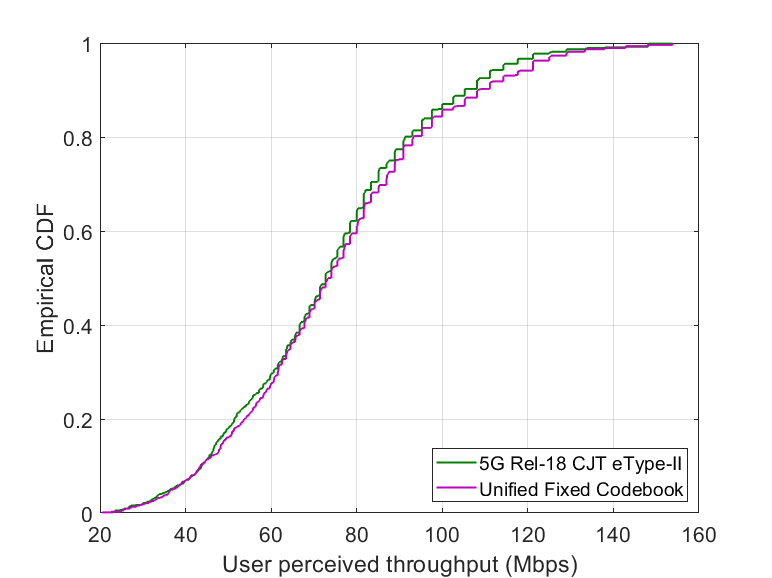}
    \caption{CDF of UPT for 5G NR CSI codebook and unified fixed codebook in the multi-TRP CJT scenario.}
    \label{fig:Cdf multi-TRP}
\end{figure}

While D-MIMO/multi-TRP architectures promise multiplicative spectral efficiency gains, realizing these benefits in practice, particularly for coherent joint transmission (CJT), can impose severe burden on the uplink resource usage. CJT requires precise cross-TRP phase and amplitude information at the network side, which increases the CSI feedback overhead.
To overcome this, the core design philosophy of the aforementioned 6G UFC, including layer-specific spatial-domain basis selection, the FD compression, the elimination of the NZC bitmap, and the flexible use of $L$ (small value of $L$) to unify low- and high-resolution feedback, can be seamlessly extended and applied to multi-TRP. By tailoring the UFC structure to capture inter-TRP phase and amplitude along with the spatial-domain basis selection potentially different for each TRP, it becomes possible to curtail the overhead associated with CJT while maintaining the performance benefit.
Figs. \ref{fig:UPT multi-TRP} and \ref{fig:Cdf multi-TRP} validate this approach by comparing the legacy 3GPP 5G Release-18 CJT eType-II codebook and the extended UFC assuming multi-TRP setup described in \cite{ref10}.

Fig. \ref{fig:UPT multi-TRP} shows the trade-off between CSI overhead and UPT gain. The 5G Release-18 CJT eType-II scheme requires over 400-bit feedback overhead to achieve a 105\% throughput gain, and scaling up to 800 bits for marginal further improvements. In contrast, the UFC demonstrates a superior overhead-efficiency trajectory. It surpasses the 105\% gain threshold utilizing fewer than 300 bits, thereby attaining aggressive feedback overhead reduction while delivering non-trivial throughput enhancement. The empirical CDFs of UPT for all users in the simulation layout, shown in Fig. \ref{fig:Cdf multi-TRP}, further support this. The CDF of the UPT throughput for the proposed 6G codebook is slightly to the right of that for the Release-18 CJT eType-II scheme. This indicates that despite the aggressive compression derived from the UFC framework, there is no degradation in the actual throughput distribution across all UEs in the layout. In other words, these findings suggest that for ensuring competitive yet affordable multi-TRP performance in 6G, the adoption of CSI codebook design that unifies single- and multi-TRP with reduced feedback overhead is essential.

\subsubsection{CSI Reference Signal}

In 5G NR, CSI-RS is used to estimate the downlink channel, enabling scheduling and beamforming operations. Meanwhile, the demodulation reference signal (DM-RS) facilitates channel estimation, supporting accurate symbol demodulation. In 6G, securing macro coverage with the anticipated adoption of FR3 and hundreds of BS antenna ports would lead to a significant increase in reference signal overhead compared to 5G. In the following, we introduce an approach to support sparse CSI-RS configurations for reducing overhead and leveraging CSI extrapolation for non-sounded frequencies~\cite{R1-2508800:sam,burghal2023enhanced}. 

\begin{figure}[h]	
	\subfloat[]{
		\includegraphics[width=0.45\linewidth]{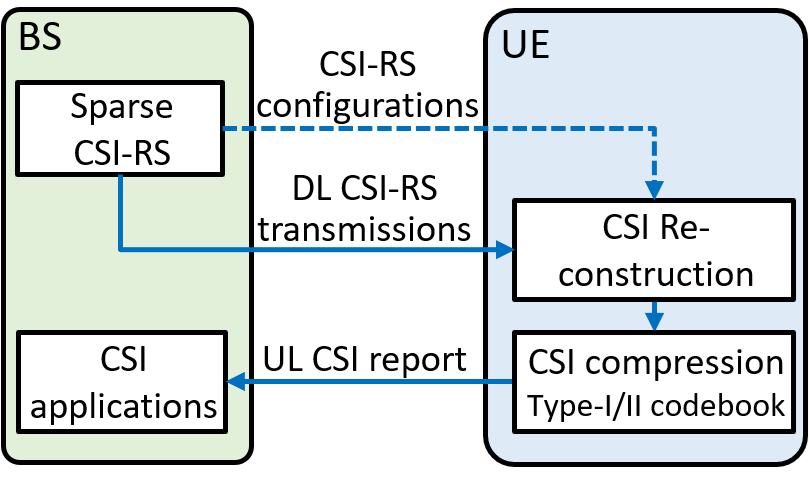}
		\label{fig:sparse_csirs_fw_a}
	}\hfill
	\subfloat[]{
		\includegraphics[width=0.48\columnwidth]{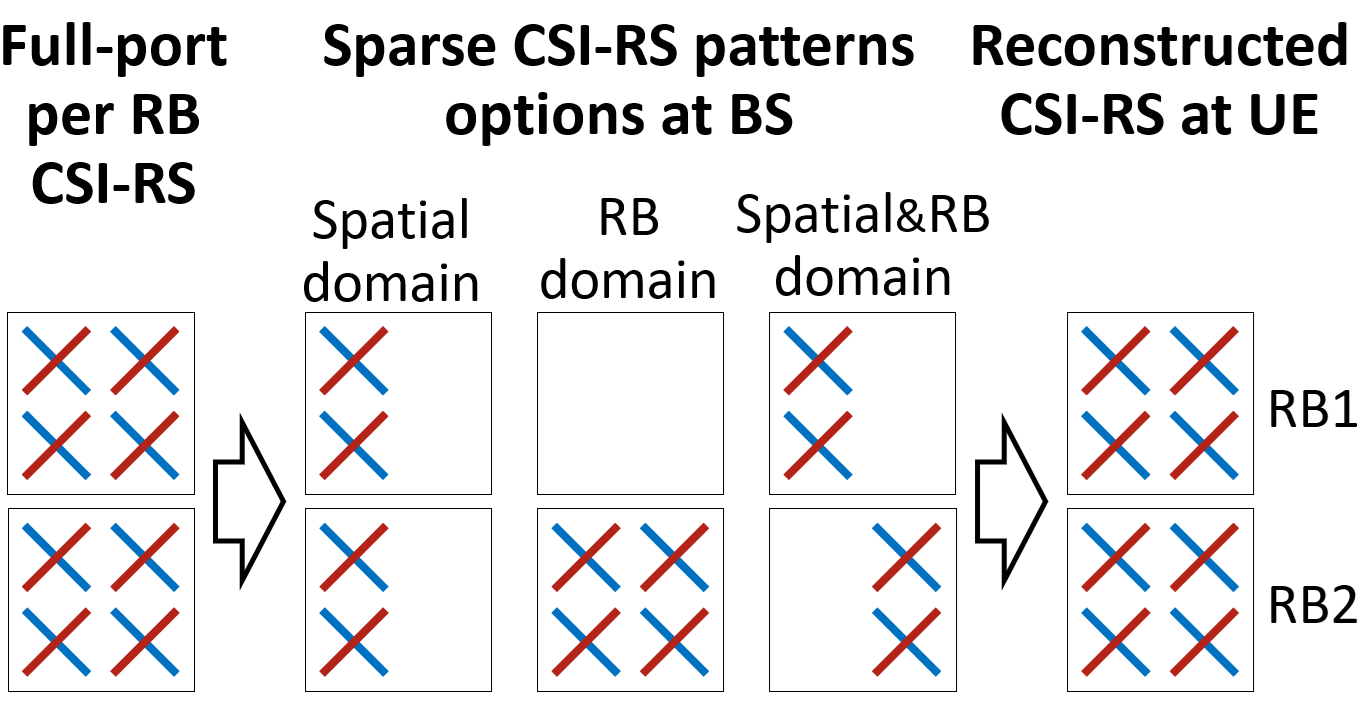}
		\label{fig:sparse_csirs_fw_b}
	}
	\caption{
		(a) Sparse CSI-RS framework. 
		(b) Sparse CSI-RS patterns and reconstruction.
	}		
\end{figure}

A key solution to address the CSI-RS overhead challenge is the adoption of sparse CSI-RS configurations. By sub-sampling CSI-RS ports across both spatial and frequency domains, sparse CSI-RS significantly reduces the resource allocation needed for each CSI-RS reception, enhancing overall system efficiency. The operational framework for sparse CSI-RS is depicted in Fig.~\ref{fig:sparse_csirs_fw_a}. This process entails the BS selecting a sparse CSI-RS pattern and signaling to the UE the specific CSI-RS ports and corresponding frequency resources for transmission. Fig.~\ref{fig:sparse_csirs_fw_b} illustrates examples of sparse CSI-RS patterns, demonstrating the selection of 50\% of ports in the spatial, frequency, or both domains. Upon receiving the sparse signal, the UE is tasked with reconstructing the full-port CSI from the available CSI-RS. Advanced techniques, such as AI-based methods, play a crucial role in ensuring the accuracy and reliability of this reconstruction process. The UE then reports the quantized CSI, for instance, using predefined codebooks, back to the BS. Within this framework, the BS can optimize overall performance by strategically selecting the frequency and spatial sparsity of the CSI-RS. Additionally, leveraging AI to learn and implement an enhanced sparse CSI-RS pattern tailored to the serving cell further improves system efficiency and reliability \cite{daoud2025pilot}.

\subsubsection{Joint Source-Channel Coding and Modulation}
    
The existing AI-based CSI compression schemes in 5G are implemented within the conventional separate source and channel coding (SSCC) pipeline, where the compressed CSI is subsequently quantized, channel-coded, and modulated as independent blocks. This block-wise separation fundamentally limits end-to-end efficiency and can lead to a performance degradation (i.e., the ``cliff effect'') once the uplink channel quality falls below the decoding threshold. 
    
To address these limitations, particularly the cliff effect, joint source-channel coding and modulation (JSCM) has emerged as a promising paradigm for 6G CSI feedback. By integrating compression and channel coding into a single end-to-end optimized model, JSCM avoids explicit quantization and mitigates the cliff effect. For CSI feedback, the deep JSCC framework~\cite{9954153} directly maps the compressed CSI to continuous modulated symbols, demonstrating clear gains over SSCC baselines. Subsequent studies have extended this paradigm to joint CSI feedback and precoding~\cite{10660530}, and digital-compatible implementations ~\cite{11563906}.  
However, applying these theoretical merits to practical cellular networks reveals severe deployment barriers. Standard two-sided JSCM solutions require matched neural networks at both the UE and BS, leading to inter-vendor model misalignment and heavy coordination overhead. Furthermore, implementing complex neural encoders on UEs significantly increases the computational and storage burden for low-power terminals. 

\begin{figure}[h]
		\centering
		\includegraphics[width=0.9\linewidth]{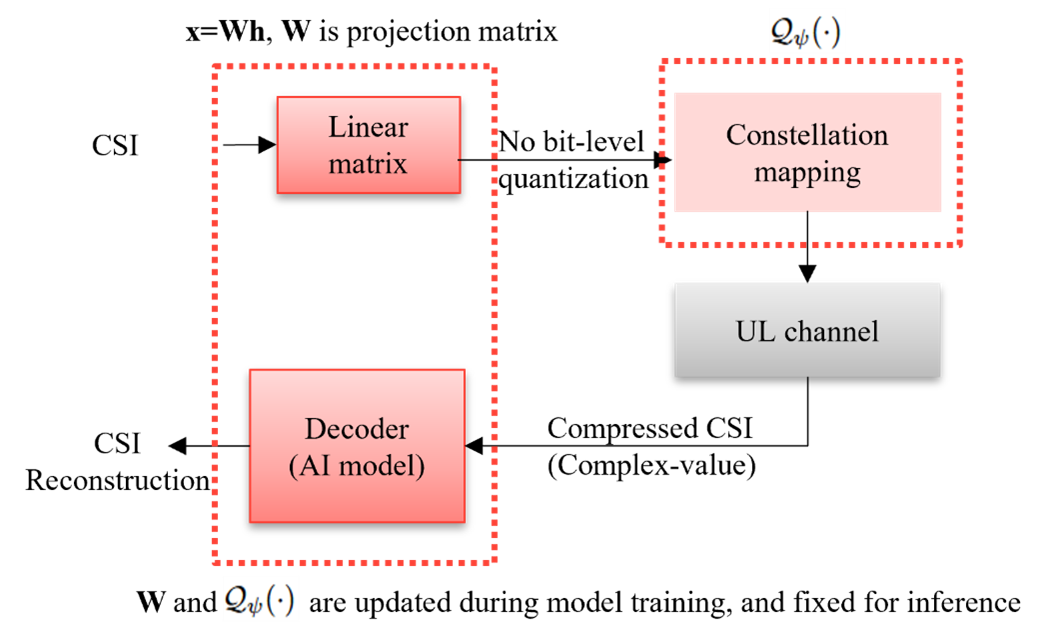}
		\caption{An example of one-sided JSCM framework for CSI compression.}
		\label{fig:JSCM}
	\end{figure}

To circumvent these multi-vendor coordination bottlenecks and terminal power constraints, one-sided JSCM designs \cite{ sun2025csi} have emerged as a key architectural solution, shifting the neural processing requirement only to the BS.
The essence of this approach is an asymmetric one-sided architecture, as depicted in Fig. \ref{fig:JSCM}. By moving all non-linear processing to the network side, the design substantially reduces terminal complexity and mitigates inter-vendor dependency in tightly coupled encoder-decoder implementations. Let $\mathbf{h} \in \mathbb{C}^{N}$ denote the downlink CSI vector. As illustrated in Fig. \ref{fig:JSCM}, the UE-side operations are simplified to a lightweight linear projection followed by a learnable constellation (modulation) mapping:

\vspace{-5pt}
\begin{equation}
\mathbf{x} = \mathbf{W}\mathbf{h}, \quad \mathbf{s} = \mathcal{Q}_{\psi}(\mathbf{x})
\vspace{-3pt}
\end{equation}
where $\mathbf{W} \in \mathbb{C}^{K \times N}$ is a spatial-frequency projection matrix and $\mathcal{Q}_{\psi}(\cdot)$ denotes a parameterized discrete constellation mapper. With this streamlined design, the UE performs only matrix multiplication and symbol mapping, while the network-side model $f_{\theta}(\cdot)$ carries out CSI reconstruction from the equalized (noisy) received symbols $\tilde{\mathbf{s}}$, i.e., $\hat{\mathbf{h}} = f_{\theta}(\tilde{\mathbf{s}})$.

\begin{figure}[h]
		\centering
		\includegraphics[width=0.9\linewidth]{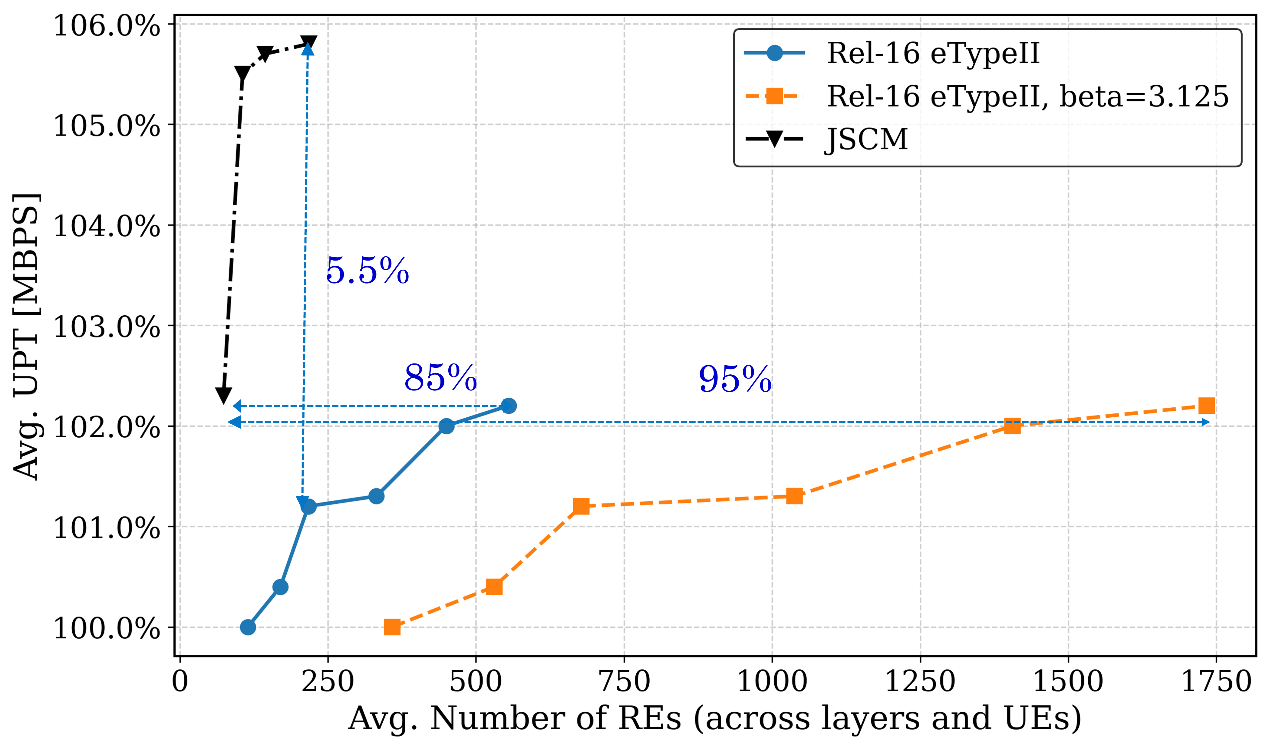}
		\caption{UPT vs. overhead for single-TRP scenario.}
		\label{fig:sTRPE}
	\end{figure}

\begin{figure}[h]
		\centering
		\includegraphics[width=0.9\linewidth]{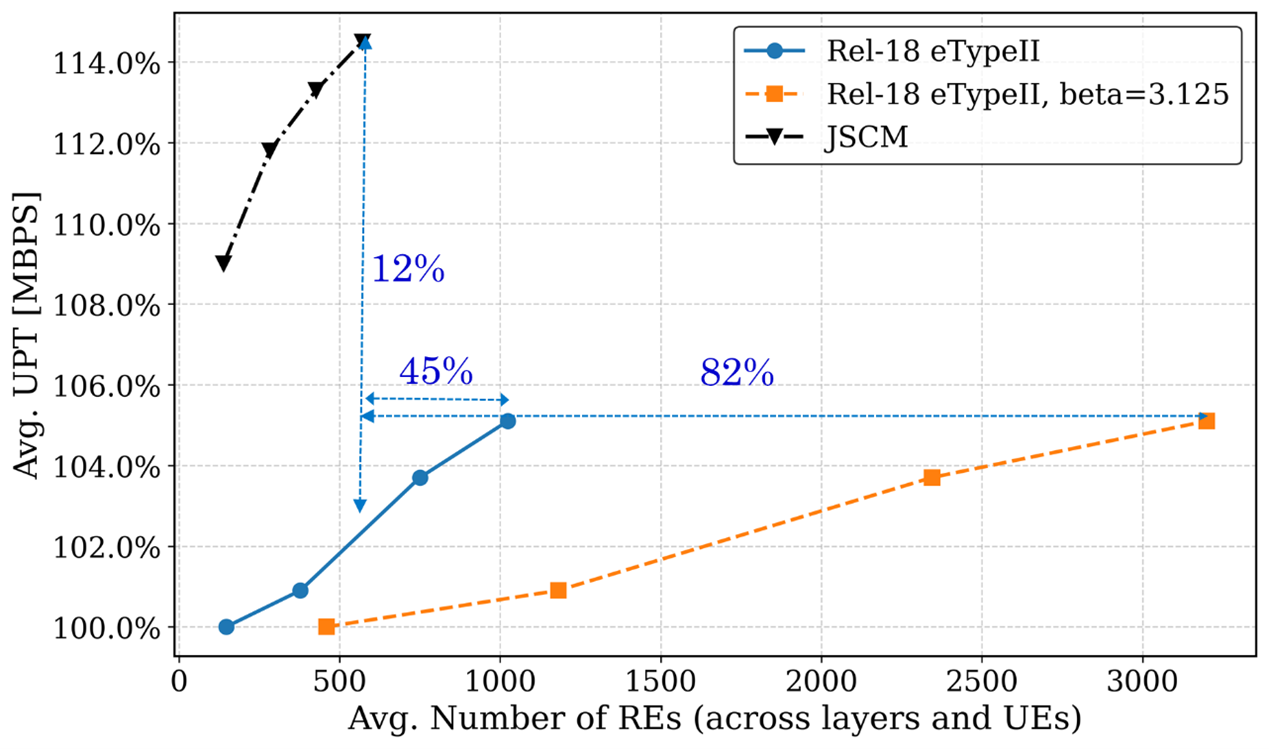}
		\caption{UPT vs. overhead for multi-TRP CJT scenario.}
		\label{fig:UPTRE}
	\end{figure}

SLS is conducted under single-TRP (\(N_{\mathrm{t}}=64\)) and multi-TRP CJT (3 TRPs × 16 ports) scenarios using FTP model 3 traffic and SU-MIMO scheduling to evaluate the aforementioned JSCM against the 5G Release-16/18 eType-II benchmarks. For eType-II codebooks, the worst-case resource element (RE) overhead piggybacked on physical uplink shared channel (PUSCH) is approximated as \(N_{\mathrm{RE}} \approx \frac{\beta \cdot B}{R_c \log_2 M_{\mathrm{mod}}}\), where $B$, \(R_{c}\), and \(M_{\mathrm{mod}}\) denote the number of CSI bits, PUSCH coding rate, and modulation order, respectively. To meet the stringent uplink control information BLER requirement, a scaling factor of $\beta$ = 3.125 is applied.
Simulation results in Figs. \ref{fig:UPTRE} and \ref{fig:sTRPE} demonstrate that the described JSCM scheme achieves a more favorable trade-off between UPT performance and overhead than the benchmarks. In the single-TRP scenario, JSCM provides a 5.5\% average UPT gain at 74 REs, and reduces feedback overhead by 85\%–95\% while maintaining 102\% of the benchmark UPT. In the multi-TRP CJT scenario, the gains are more pronounced, yielding a 12\% UPT improvement alongside a 45\%–82\% overhead reduction. The improvement stems from the fact that JSCM's joint optimization effectively captures the highly irregular channel characteristics of multi-TRP environments, where fixed codebooks fail to quantize efficiently.

\subsection{Neural Receivers}
\label{neuralreceivers}

While the previous section discusses AI/ML for CSI, that topic can also be viewed as one example of a much broader transformation now emerging in wireless system design. CSI acquisition, compression, and prediction are among the earliest physical-layer functions where AI/ML has shown practical promise, but they are part of a larger trend in which traditionally model-based signal processing blocks are being revisited through a data-driven lens. Recent advances in AI/ML have moved well beyond image recognition and natural language processing and are now being aggressively explored for wireless communications, including 5G-Advanced and prospective 6G \cite{hoydis2021toward}. AI/ML adoption is reopening many assumptions that underpinned the traditional air-interface work, indicating that the PHY layer is evolving in ways that are not yet settled \cite{o2017introduction, qin2019deep}. Many real-world phenomena, such as nonlinear RF behavior and sophisticated interference, are not well captured by classical communication models. AI/ML models can learn effective representations and nonlinear mappings directly from data, achieving better fidelity and robustness than hand-crafted approximations. This is applicable to PHY layer tasks such as channel estimation, equalization, and impairment compensation.

The current 5G-Advanced standardization work provides evidence of where AI/ML is entering the PHY layer and how much design flexibility still exists. AI/ML techniques have been evaluated and standardized for tasks such as CSI feedback (including CSI compression and CSI prediction), beam management, and positioning \cite{lin2023embracing}. These AI/ML-assisted functions can be viewed as precursors to a more radical evolution: the PHY layer is being transformed from a fully specified algorithmic pipeline into a data-driven, continuous-learning subsystem. Unlike conventional PHY layer designs, AI/ML techniques can be used to treat key baseband functions as learnable entities subject to continuous optimization under site-specific conditions. This evolution challenges the notion of a ``mature'' PHY layer in the classical sense.

\subsubsection{From Model-based to Hybrid and Data-driven Architectures}

Classical PHY layer design is fundamentally model-based. One assumes a parametric channel model, $p(\boldsymbol{y} | \boldsymbol{x}; \boldsymbol{\phi})$, which is a probabilistic description of how the channel turns transmitted signal $\boldsymbol{x}$ into received signal $\boldsymbol{y}$ under the model parameterized by $\boldsymbol{\phi}$ capturing effects such as large- and small-scale fading, and white Gaussian noise. Waveforms, modulations, coding schemes, channel estimators, equalizers, and detectors are then individually optimized under this model. The design typically targets closed-form solutions and optimization objectives such as capacity, mutual information, or minimum mean-squared error (MMSE). This paradigm has been remarkably successful, but it becomes sub-optimal in regimes where the assumed model is an inaccurate approximation of reality, including, e.g., non-stationarity, hardware non-idealities, and near-field effects. In such environments, the modeling gap, i.e., the difference between the assumed model $p(\boldsymbol{y} | \boldsymbol{x}; \boldsymbol{\phi})$ and the true physical process, can dominate performance. Put differently, the bottleneck is not lack of optimality within the model but the inadequacy of the model itself.

AI/ML-based designs for the PHY layer address the gap by shifting from purely model-based optimization to hybrid or even fully data-driven approaches \cite{zappone2019wireless}. The central idea is to retain analytical structure where models perform well and to replace or augment sub-optimal components with learnable functions trained on representative data. Denote by $\boldsymbol{\theta}$ the parameters of one or more learnable PHY layer components, and denote by $\mathcal{D}$ a dataset collected under deployment or high-fidelity simulation conditions. Rather than optimizing under an assumed parametric channel model $p(\boldsymbol{y} | \boldsymbol{x}; \boldsymbol{\phi})$, the effective design problem becomes:
\begin{equation}
\boldsymbol{\theta}^\star = \arg \min_{\boldsymbol{\theta}} \mathbb{E}_{(\boldsymbol{x}, \boldsymbol{y}) \sim \mathcal{D} } [\mathcal{L} ( \boldsymbol{x}, \hat{\boldsymbol{x}} (\boldsymbol{y}; \boldsymbol{\theta}) ) ],
\end{equation}
where $\mathcal{L} (\cdot)$ denotes the loss function for performance optimization and $\hat{\boldsymbol{x}} (\boldsymbol{y}; \boldsymbol{\theta}) $ denotes the estimate of the ground truth $\boldsymbol{x}$ under the observation $\boldsymbol{y}$ and the learnable parameters $\boldsymbol{\theta}$. The architecture retains models that perform well, while learning is used to close gaps where models are inaccurate. As a result, achievable performance becomes a function of both physical resources (e.g., bandwidth and power) and data availability/quality, a departure from the classical, purely model-driven notion of maturity.

\subsubsection{Neural Receivers as Canonical AI/ML-based PHY Layer Components}

Neural receivers are a canonical embodiment of AI/ML based PHY layer design. In contrast to classical architectures that compose fixed channel estimation, equalization, and demapping blocks, a neural receiver treats reception as a joint learning problem: map what the receiver observes over time-frequency-space directly to soft bits \cite{honkala2021deeprx, cammerer2023neural}. Specifically, denote by $\boldsymbol{Y}$ the received resource grid (e.g., across antennas, subcarriers, and OFDM symbols), and denote by $\boldsymbol{p}$ the associated reference signals. A neural receiver implements a parameterized mapping $f(\cdot)$:
\begin{equation}
\hat{\boldsymbol{\ell}} = f (\boldsymbol{Y}, \boldsymbol{p}; \boldsymbol{\theta}), 
\end{equation}
where $\hat{\boldsymbol{\ell}}$ are bit LLRs and $\boldsymbol{\theta}$ are neural network parameters. Training typically minimizes binary cross-entropy $\mathcal{L} ( \boldsymbol{b}, \hat{\boldsymbol{\ell}})$ between coded bits $\boldsymbol{b}$ and LLRs $\hat{\boldsymbol{\ell}}$.

\begin{figure}[!t]
\centering
\includegraphics[width=3.7in]{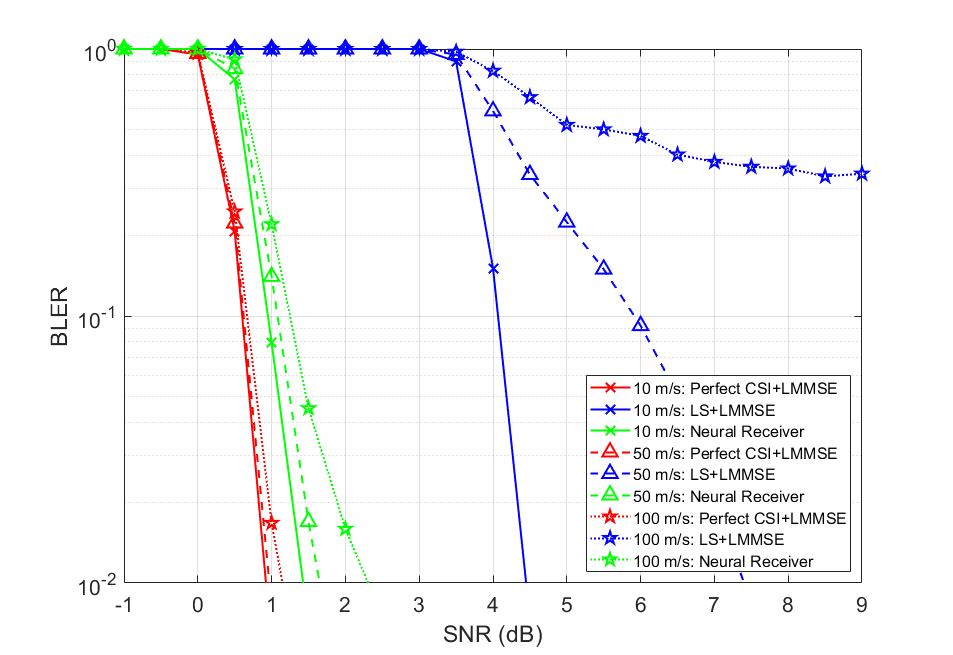}
\caption{BLER performance of a neural receiver under different speeds and 16-QAM.}
\label{fig:nrx-comb}
\end{figure}

Because the neural network mapping $f(\cdot)$ is trained on specific propagation conditions, neural receivers can learn channel propagation effects, including temporal, frequency, and spatial correlations and hardware-induced distortions. This enables them to recover bits with improved BLER performance. As an example, Fig. \ref{fig:nrx-comb} shows the BLER performance of a neural receiver for joint channel estimation, equalization, and demapping, across three UE speeds (10 m/s, 50 m/s, and 100 m/s) and 16-QAM  \cite{sionna}. Two baselines are considered: 1) linear MMSE (LMMSE) equalization with perfect CSI, and 2) LMMSE equalization with least square (LS) estimated CSI. In Fig. \ref{fig:nrx-comb}, the perfect‐CSI curves virtually coincide for all speeds, confirming that an LMMSE equalizer with perfect CSI is agnostic to Doppler. By contrast, the LS+LMMSE curves shift dramatically rightward as speed increases, with the SNR required for 10\% BLER rising from about 4.1 dB at 10 m/s to 5.9 dB at 50 m/s and an error floor exists at 100 m/s. The neural receiver, however, maintains a much tighter proximity to the perfect-CSI benchmark. While the LS-based estimator's performance degrades steeply with Doppler, the neural model's SNR requirement increases by less than 0.3 dB across a tenfold speed range. These results highlight the neural receiver's ability to learn and compensate for rapid channel variations, delivering robust BLER performance under fast-fading scenarios.

In a broad sense, neural receivers encompass any receiver design in which one or more processing blocks are enhanced or replaced by neural networks. This admits a natural four-level taxonomy that helps to connect incremental deployments in 5G-Advanced to more radical designs in the 6G era.
\begin{itemize}
\item \textit{Level 1: Block-level enhancement.} A single processing block (e.g., channel estimation) in the receiver is substituted or augmented by a neural network model. Examples include: 1) a neural network that refines LS or LMMSE pilot-based channel estimates; and 2) a neural network that directly maps time-frequency data observations and pilots to channel estimates. The interfaces between the neural network block and other blocks are preserved, making Level 1 the most straightforward to adopt.

\item \textit{Level 2: Multi-block modular replacement.} Two or more blocks in the receiver are replaced by distinct neural network models, e.g., separate models for channel estimation, equalization, and symbol demapping. This enables incremental integration of learning while retaining a modular structure and legacy interfaces between blocks. 

\item \textit{Level 3: Joint-block integration.} A single neural network model performs multiple roles jointly, such as channel estimation, equalization, and symbol demapping. By operating across block boundaries, Level-3 neural receivers can exploit cross-block correlations (e.g., between pilot and data REs) and thus realize joint processing gains.

\item \textit{Level 4: Monolithic neural receiver chain.} At the most aggressive integration level, a monolithic neural network model approximates the entire receiver chain. In principle, this has the potential to approach optimal performance by jointly optimizing all processing stages. In practice, Level-4 designs face substantial challenges in interpretability, verification, interoperability, and robust operation under distribution shifts.

\end{itemize}

This four-level taxonomy clarifies that neural receivers are not a single architecture but a family of designs. Despite promising results, neural receivers face challenges. One key unsettled issue is robustness. Once deployed, a neural receiver encounters conditions that may differ from its training distribution. While classical receivers admit performance analyses under specified communication models, the generalization behavior of neural network models under the distribution shifts is much harder to characterize. Thus, neural receivers are best viewed as evidence that core receiver design is still evolving. They exemplify an AI/ML-based air interface design phase in which the PHY layer is increasingly shaped by learnable components and data, rather than solely by fixed analytical models.

\subsubsection{Neural Receiver-pilot Co-design}

In conventional PHY layer design, pilot patterns (e.g., DMRS in 5G) are chosen first, typically from a set of standardized options, based on assumed channel models and target Doppler/delay spreads. The receiver is then engineered, often as an LS or LMMSE estimator followed by equalization and symbol demapping, to operate under that fixed pilot structure. In this model-based paradigm, pilot design and receiver design are largely decoupled. Neural receivers can achieve more than improving receiver performance; they change what is feasible on the pilot side. In particular, the placement, density, and even the structure of pilots can be optimized jointly with neural receivers. Two complementary directions are of particular interest: 1) reduced-overhead orthogonal pilots, where neural receiver performance gains are traded for lower pilot density; and 2) non-orthogonal pilots, most notably superimposed pilots \cite{ait2021end}, where pilots and data share the same REs and a neural receiver learns to separate and exploit both.

In the classical regime, the channel impulse response $h(\tau)$ is assumed essentially supported on $[0, \tau_{\max}]$, where $\tau_{\max}$ captures the maximum channel tap delay containing most of the channel energy. Under this assumption, uniform pilot spacing in frequency at density $\tau_{\max}$ samples/Hz is sufficient to recover the channel frequency response $H(f)$, according to the Nyquist sampling theory. Therefore, 5G DMRS with regular comb-2 pilot pattern may be viewed as Nyquist sampling of the channel. A neural receiver can learn to exploit longer-range correlations in time and frequency, reconstructing the effective channel from a sparser comb of pilot REs. The BLER performance gain that the neural receiver provides at a given DMRS density can be traded for reducing pilot density while maintaining the same BLER performance. In practical terms, patterns that would previously have required comb-2 spacing can be relaxed to, e.g., comb-4, comb-8, or comb-16, freeing a fraction of REs for data.

In realistic multipath channels, the energy of the channel impulse response $h(\tau)$ is typically sparse in delay, concentrated in a small number of clusters rather than uniformly distributed on $[0, \tau_{\max}]$. Suppose the impulse response is contained in a union of disjoint intervals whose total length is $\tau_{\textrm{tot}} \ll \tau_{\max}$. If the locations of these intervals are known, Landau sampling theory shows that it is sufficient to sample in frequency at an average density $\tau_{\textrm{tot}}$ samples/Hz. In practice, the exact locations of the delay clusters are not known a priori and may change with environment. In this case, blind multi-coset sampling shows that one can still recover the channel by sampling in frequency at twice the Landau sampling rate \cite{delfeld2025sparse}. In pilot terms, this means that we can afford to use fewer pilots on average if we design their positions to reflect the sparse support structure. The pilot pattern is no longer a uniform comb; it becomes a sparser, possibly irregular pattern. A neural receiver can be trained to map from the irregularly sampled pilots plus data REs to soft bits. In effect, the neural receiver acts as a powerful non-linear interpolator/reconstructor that implicitly leverages sparsity, without requiring an explicit sparse-recovery algorithm for the underlying process. The practical implication is that dense regular pilots can be replaced by sparser, possibly irregular pilots, converting neural receiver performance gains into pilot-overhead reduction and higher net throughput.

The second direction goes further by relaxing not only density (either regular comb or irregular patterns) but also orthogonality. Instead of reserving separate REs for pilots, superimposed pilots embed pilot sequences on top of the data. Specifically, the symbol transmitted on a given resource element with subcarrier index $k$ and OFDM symbol index $n$ is given by:
\begin{equation}
x_{k,n} = \sqrt{\alpha_{k,n}} p_{k,n} + \sqrt{1 - \alpha_{k,n}} s_{k,n},
\end{equation}
where $p_{k,n}$ is the pilot symbol, $s_{k,n}$ is the data symbol, and $\alpha_{k,n} \in [0, 1]$ controls the pilot-data power split. In classical receivers, superposition is hard to exploit because pilot and data interfere with each other and simple linear estimators struggle to decouple them without sacrificing performance. In contrast, neural receivers  can learn nonlinear separation of the pilot and data components, exploiting structure in both $p_{k,n}$ and $s_{k,n}$, as well as temporal/frequency correlations in channel impulse response. The power split $\alpha_{k,n}$ can be included in the training loop, allowing joint optimization of superimposed pilot power and the neural receiver mapping.

\begin{figure}[!t]
\centering
\includegraphics[width=3.7in]{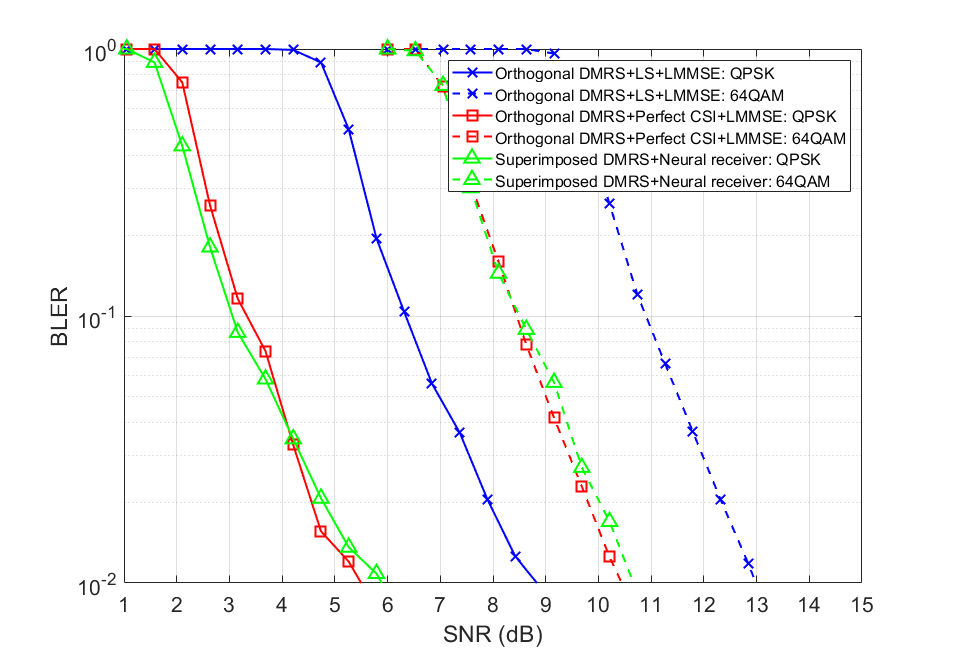}
\caption{BLER performance of superimposed pilot + neural receiver under different QAM orders and 3 m/s speed.}
\label{fig:nrx-sip}
\end{figure}

Because superimposed designs eliminate dedicated pilot REs, all REs previously reserved for pilots  become available for data, and the neural receiver's ability to separate and exploit the superimposed structure is what determines whether this reclaimed spectrum can be used without degrading reliability. As an example, Fig. \ref{fig:nrx-sip} shows the BLER performance of superimposed pilot + a neural receiver for joint channel estimation, equalization, and demapping \cite{sionna}. Two baselines are considered: 1) LMMSE equalization with perfect CSI, and 2) LMMSE equalization with LS-estimated CSI. For QPSK, Fig. \ref{fig:nrx-sip} shows that the orthogonal DMRS + perfect CSI + LMMSE curve reaches 10\% BLER around 3.3 dB. The superimposed DMRS + neural receiver tracks it closely, slightly outperforming it by ~0.2 dB at 10\% BLER. In contrast, the orthogonal DMRS + LS + LMMSE needs much higher SNR, with 10\% BLER only near ~6.3 dB. For 64-QAM, Fig. \ref{fig:nrx-sip} shows that the orthogonal DMRS + perfect CSI + LMMSE curve achieves 10\% BLER around 8.4 dB. The superimposed DMRS + neural receiver curve remains close, showing that neural receiver based pilot cancellation and estimation continue to work well even with higher-order modulation. In contrast, the orthogonal DMRS + LS + LMMSE requires roughly ~10.9 dB to reach 10\% BLER.

In short, pilot design, which used to be a fixed, model-driven part of the air interface, is now shifting toward a tunable, data-driven degree of freedom co-optimized with the neural receiver. It reopens a design space, where classical signaling, sampling theory, and learning-based inference are being recombined in fundamentally new ways. A natural question, after considering orthogonal and superimposed pilot designs enhanced by neural receivers, is whether explicit pilots are needed at all. In both cases, dedicated reference symbols remain an explicit design element around which the receiver is structured. Neural receivers, however, are not inherently tied to this paradigm. Once the receiver is capable of learning channel propagation effects and hardware impairments, it becomes conceivable to embed reference information implicitly in the data signal, as discussed in detail in the next subsection.

\subsubsection{Pilot-free Transmission Enabled by Neural Receivers}

Pilot-free (e.g., DMRS-free) transmission represents a radical direction enabled by neural receivers. Instead of transmitting explicit reference signal for channel estimation, the transceiver is designed so that the data signal itself carries sufficient structure for the receiver to infer the channel and recover data by leveraging learned mappings rather than explicit model-based estimators. 

Specifically, consider a transmitter that implements a mapping function $f_{\textrm{tx}} (\cdot)$ as:
\begin{equation}
\boldsymbol{x} = f_{\textrm{tx}} (\boldsymbol{b}; \boldsymbol{\theta}_{\textrm{tx}}),
\end{equation}
where $\boldsymbol{x}$ denotes the transmitted signal, $\boldsymbol{b}$ denotes the coded bits, and $\boldsymbol{\theta}_{\textrm{tx}}$ denotes learnable parameters (including, for example, a learned constellation) at the transmitter side. The received signal at the receiver is given by:
\begin{equation}
\boldsymbol{Y} = f_{\textrm{ch}} (\boldsymbol{x}) + \boldsymbol{N},
\end{equation}
where the mapping function $f_{\textrm{ch}} (\cdot)$ captures propagation and hardware impairments, and $\boldsymbol{N}$  denotes noise and interference. In a pilot-free design, there are no explicit pilot resources; all scheduled REs carry data generated via $f_{\textrm{tx}} (\cdot)$. A neural receiver $f_{\textrm{rx}} (\cdot)$ with learnable parameters $\boldsymbol{\theta}_{\textrm{rx}}$ maps the received signal $\boldsymbol{Y}$ directly to soft bits. The joint design problem can be formulated as:
\begin{equation}
(\boldsymbol{\theta_{\textrm{tx}}^\star}, \boldsymbol{\theta_{\textrm{rx}}^\star}) = \arg \min_{\boldsymbol{\theta_{\textrm{tx}}}, \boldsymbol{\theta_{\textrm{rx}}}} \mathbb{E}_{(\boldsymbol{b}, \boldsymbol{Y}) \sim \mathcal{D} } [\mathcal{L} ( \boldsymbol{b}, f_{\textrm{rx}} (\boldsymbol{Y}; \boldsymbol{\theta}_{\textrm{rx}}) ) ],
\end{equation}
subject to transmitter constraints on $\boldsymbol{x}$ such as peak and average power constraints.

\begin{figure}[!t]
\centering
\includegraphics[width=3.7in]{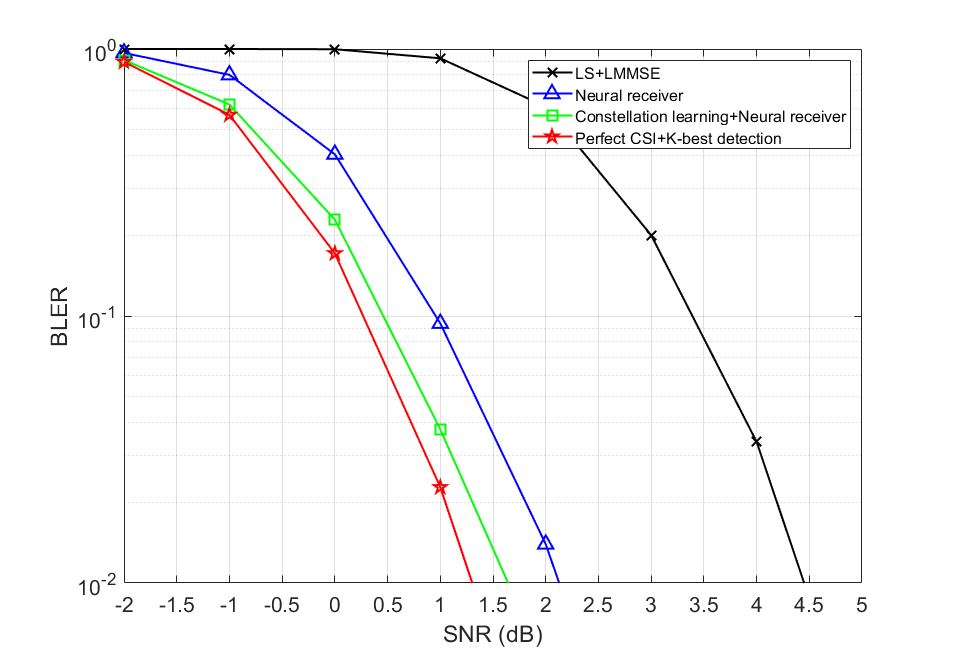}
\caption{BLER performance of pilot-free transmission with a neural receiver under 16-QAM.}
\label{fig:nrx-pilotfree}
\end{figure}

A central enabler of pilot-free transmission is the use of learned constellations that make the joint inference of data and channel feasible \cite{ait2021end}. Instead of fixed QAM, the encoder can learn a constellation (geometry and labeling) that is robust to channel uncertainty and facilitates inference. For example, constellation points may be arranged to maximize distinguishability under a site-specific channel distortion. Certain patterns in the learned constellation effectively act as implicit pilots without occupying explicit REs. In this sense, pilot-free transmission does not eliminate the need for reference information; it learns to embed such information in the data signal itself, in ways that classical design does not typically consider. As an example, Fig. \ref{fig:nrx-pilotfree} shows the BLER performance of the pilot-free transmission with a neural receiver under 16-QAM \cite{sionna}. Two baselines are considered: 1) K-best detection with perfect CSI, and 2) LMMSE equalization with LS-estimated CSI. In Fig. \ref{fig:nrx-pilotfree}, LS channel estimation + LMMSE equalization with regular QAM performs the worst: it achieves 10\% BLER around 3.4 dB. Neural receiver with regular QAM is a large step up: it reaches 10\% BLER near 1.0 dB, about 2.4 dB better than LS channel estimation + LMMSE equalization. Constellation-learning + neural receiver adds another 0.5 dB gain at 10\% BLER vs. neural receiver only. Perfect CSI + K-best detection with regular QAM is the reference bound. It sits a further ~0.2 dB to the left of the curve corresponding to constellation-learning + neural receiver. The key takeaway is that pilot-free transmission with a neural receiver is within ~0.2 dB of the perfect-CSI + K-best detection bound, implying that pilot-free transmission with a neural receiver can approach near-optimal detection with perfect CSI.

Despite promising potential, pilot-free transmission intensifies several open challenges. Training relies on supervised learning from known bit sequences and their corresponding received signals, without explicit channel labels. Ensuring that the learned mapping remains identifiable and well-conditioned across the intended range of channel conditions (e.g., SNR, delay/Doppler spreads, hardware profiles) is non-trivial. Insufficient coverage in $\mathcal{D}$ may lead to overfitting. Distribution shifts can be detrimental when the receiver must jointly infer data and channel from data-only observations. Nonetheless, pilot-free transmission with neural receivers revisits a foundational design element, i.e., the use of explicit pilots, which has been largely regarded as fixed in modern communication systems. It demonstrates that, when transceiver design is reformulated as a joint learning problem, previously rigid overheads (e.g., DMRS patterns) can be relaxed, uncovering additional spectral efficiency gains and new design degrees of freedom.

In short, the rise of AI/ML-based design for the PHY layer challenges any simplistic claim of closure. The incorporation of learning fundamentally alters the design landscape. It couples PHY layer performance to data and compute constraints, so that achievable reliability and throughput now depend as much on dataset coverage and accelerator capability as on SNR and bandwidth \cite{kundu2025ai}. It demands new theoretical frameworks for performance guarantees, new operational processes for training, validation, and life cycle management of models, and new standardization principles for specifying learnable components and their interfaces rather than fully prescribing algorithms. In this sense, rather than signaling maturity, AI/ML-based design marks a transition into a new phase of PHY layer innovation in which adaptability to site-specific conditions emerges as a core design principle.

\section{Conclusion}
\label{sec:conclusion}

\begin{table*}[h!]
\centering
%\begin{tabular}{|c|c|c|}
\begin{tabular}{|p{5cm}|p{5cm}|p{5cm}|}
\hline
\textbf{Potential improvement areas} & \textbf{Performance gains under respective simulation setups} & \textbf{Complexity} \\ \hline

Constellation Shaping & 1.0--1.4 dB SNR gain &   Maximum shaping gain is only realizable for higher order constellations ($\ge256$-QAM) with large DM size ($\ge 1024$ systems) and at high SNR\\ \hline

New LDPC Codes & 0.55 dB SNR gain & Similar complexity as 5G BG1 \\ \hline

Reduced-complexity UE receivers & $\ge$ 2.0--3.5dB with PS & Receiver complexity is one third to one tenth of the conventional sphere decoder \\ \hline

Waveforms & Nominal 5-10 \% spectral efficiency gains in certain scenarios & Receiver complexity is higher relative to CP-OFDM  \\ \hline

MIMO and CSI design & 
- 95 percentile UPT: SISO $\approx{}36 \textrm{Mbps}$; $\textrm{SU-MIMO}$ $\approx{}78 \textrm{Mbps}$; $\textrm{sum MU-MIMO}$ $\approx{}120 \textrm{Mbps}$; $\textrm{sum D-MIMO}$ $\approx{}150 \textrm{Mbps}$

- Unified fixed codebook: 13\% average UPT gain for single-TRP, and 5\% average UPT gain for multi-TRP CJT

- JSCM: 5.5\% average UPT gain for single-TRP and 12\% average UPT gain for multi-TRP CJT

 & Unified fixed codebook: Low-complexity implementation achieved by layer-specific spatial-domain basis selection and eliminating NZC bitmaps.

JSCM: Low UE complexity by integrating source coding, channel coding, and modulation into a single simplified step (linear projection and RE mapping). \\ \hline

Neural Receivers& 2–3 dB SNR gain vs. LS/LMMSE in pilot-limited settings & Receiver complexity is higher relative to LS/LMMSE \\ \hline

\end{tabular}
\caption{Potential improvement areas in the PHY layer and their complexity to realize.}
\label{tab:areasofgains}
\end{table*}

This paper has assessed whether the PHY layer is approaching maturity. Table~\ref{tab:areasofgains} summarizes the potential gains and realization complexity of the techniques considered. Constellation shaping, channel-code refinements, reduced-complexity receivers, and waveform evolution remain important, but their gains are generally bounded and must be balanced against overhead and implementation complexity. Thus, for many conventional link level techniques, further spectral efficiency gains are entering a regime of diminishing returns. The largest scalable opportunity remains the spatial domain. Massive MIMO can provide array, interference-suppression, and spatial-multiplexing gains when CSI can be acquired efficiently through TDD reciprocity. In practice, RF calibration, pilot reuse, channel aging, and weak pilot reception from power-limited cell-edge UEs can limit CSI accuracy and hence the achievable gain. Distributed MIMO can further improve macro diversity and interference management, but coherent operation requires adequate fronthaul, synchronization, calibration, CSI exchange, and scalable coordination. AI/ML provides a complementary source of innovation for channel estimation, detection, CSI compression, and adaptation. Overall, the PHY layer is not dead: its most consequential advances will come from scalable spatial processing and deployable CSI acquisition, complemented by targeted link-level refinements and AI/ML-based designs.

\section*{Acknowledgment}

The authors express their gratitude to Professor Erik G. Larsson for many useful discussions regarding this manuscript. We also thank Dr. Harsh Tataria for his valuable comments.

\ifCLASSOPTIONcaptionsoff
  \newpage
\fi

\bibliographystyle{IEEEtran}
\bibliography{references.bib}

@book{TSE,
  author        = "D. Tse and P. Visawanath",
  title         = "Fundamentals of Wireless Communications",
  publisher     = "Cambridge",
  address       = "New York, NY",
  year          = "2005"
}

@misc{shafi2025industrialviewpointsrantechnologies,
      title={Industrial Viewpoints on {RAN} Technologies for {6G}}, 
      author={Mansoor Shafi and Erik G. Larsson and Xingqin Lin and Dorin Panaitopol and Stefan Parkvall and Flavien Ronteix-Jacquet and Antti Toskala},
      year={2025},
      eprint={2508.08225},
      archivePrefix={arXiv},
      primaryClass={cs.NI},
      url={https://arxiv.org/abs/2508.08225}, 
}

@TECHREPORT {ITURM2410,
    author      = {ITU-R},
    title       = {{ITU-R M2410} - Minimum requirements related to  
technical performance for {IMT-2020}  
radio interface(s)},
    year    = {2017},
    institution= {ITU-R},
     }

@ARTICLE{775355,
  author={Alouini, M.-S. and Goldsmith, A.J.},
  journal={IEEE Transactions on Vehicular Technology}, 
  title={Area spectral efficiency of cellular mobile radio systems}, 
  year={1999},
  volume={48},
  number={4},
  pages={1047-1066},
  doi={10.1109/25.775355}}

@ARTICLE{6730658,
  author={Rysavy, Peter},
  journal={Proceedings of the IEEE}, 
  title={Challenges and Considerations in Defining Spectrum Efficiency}, 
  year={2014},
  volume={102},
  number={3},
  pages={386-392},
  doi={10.1109/JPROC.2014.2301637}}

@ARTICLE{3GPPTS38306,
  author={3GPP},
  title={User Equipment ({UE}) radio access capabilities (v.19.1.0)}, 
  journal={{TS 38.306}},
  year={2026}}

@ARTICLE{1091871,
  author={Massey, J.},
  journal={IEEE Communications Magazine}, 
  title={Information theory: The copernican system of communications}, 
  year={1984},
  volume={22},
  number={12},
  pages={26-28},
  doi={10.1109/MCOM.1984.1091871}}

@INPROCEEDINGS{turbo,
  author={Berrou, C. and Glavieux, A. and Thitimajshima, P.},
  booktitle={Proceedings of ICC '93 - IEEE International Conference on Communications}, 
  title={Near {Shannon} limit error-correcting coding and decoding: Turbo-codes. 1}, 
  year={1993},
  volume={2},
  number={},
  pages={1064-1070 vol.2},
  doi={10.1109/ICC.1993.397441}}

@techreport{6GWS-25243,
    author = {3GPP},
    title = {Chair's summary of the {3GPP} workshop on {6G}},
    institution = {{3GPP}},
    year = 2025,
    month = mar,
    number = {{6GWS-25243}},
    address = {Seoul, Korea}
}

@ARTICLE{1092061,
  author={Foschini, G. and Gitlin, R. and Weinstein, S.},
  journal={IEEE Transactions on Communications}, 
  title={Optimization of Two-Dimensional Signal Constellations in the Presence of {Gaussian} Noise}, 
  year={1974},
  volume={22},
  number={1},
  pages={28-38},
  doi={10.1109/TCOM.1974}}

@INPROCEEDINGS{11154592,
  author={Ivanov, Kirill and Yang, Wei and Jiang, Jing},
  booktitle={2025 13th International Symposium on Topics in Coding (ISTC)}, 
  title={Probabilistic Shaping in {MIMO}: Going Beyond 1.53 dB {AWGN} Gain With the Non-Linear Demapper}, 
  year={2025},
  volume={},
  number={},
  pages={1-5},
  doi={10.1109/ISTC65386.2025.11154592}}

@article{Li_2011,
   title={Reduced complexity sphere decoding},
   volume={11},
   ISSN={1530-8677},
   url={http://dx.doi.org/10.1002/wcm.1216},
   DOI={10.1002/wcm.1216},
   number={12},
   journal={Wireless Communications and Mobile Computing},
   publisher={Wiley},
   author={Li, Boyu and Ayanoglu, Ender},
   year={2011},
   month=nov, pages={1518–1527} }

@ARTICLE{4444760,
  author={Studer, Christoph and Burg, Andreas and Bolcskei, Helmut},
  journal={IEEE Journal on Selected Areas in Communications}, 
  title={Soft-output sphere decoding: algorithms and {VLSI} implementation}, 
  year={2008},
  volume={26},
  number={2},
  pages={290-300},
  doi={10.1109/JSAC.2008.080206}}

@ARTICLE{1408197,
  author={Jalden, J. and Ottersten, B.},
  journal={IEEE Transactions on Signal Processing}, 
  title={On the complexity of sphere decoding in digital communications}, 
  year={2005},
  volume={53},
  number={4},
  pages={1474-1484},
  doi={10.1109/TSP.2005.843746}}

@ARTICLE{6736761,

  author={Larsson, Erik G. and Edfors, Ove and Tufvesson, Fredrik and Marzetta, Thomas L.},

  journal={IEEE Communications Magazine}, 

  title={Massive MIMO for next generation wireless systems}, 

  year={2014},

  volume={52},

  number={2},

  pages={186-195},

  doi={10.1109/MCOM.2014.6736761}}

@misc{boccuzzi2025spectralefficiencyconsiderations6g,
      title={Spectral Efficiency Considerations for {6G}}, 
      author={Joseph Boccuzzi},
      year={2025},
      eprint={2508.09117},
      archivePrefix={arXiv},
      primaryClass={eess.SP},
      url={https://arxiv.org/abs/2508.09117}, 
}

@ARTICLE{1459002,
  author={Burg, A. and Borgmann, M. and Wenk, M. and Zellweger, M. and Fichtner, W. and Bolcskei, H.},
  journal={IEEE Journal of Solid-State Circuits}, 
  title={{VLSI} implementation of {MIMO} detection using the sphere decoding algorithm}, 
  year={2005},
  volume={40},
  number={7},
  pages={1566-1577},
  doi={10.1109/JSSC.2005.847505}}

@techreport{qcom,
    author = {Qualcomm},
    title = {Modulation, joint channel coding and modulation for {6GR}},
    institution = {{3GPP}},
    year = 2025,
    month = Aug,
    number = {{R1-2506221}},
    address = {Banglore, India}}

@ARTICLE{sixchallenges,
  author={Tataria, Harsh and Shafi, Mansoor and Dohler, Mischa and Sun, Shu},
  journal={IEEE Vehicular Technology Magazine}, 
  title={Six Critical Challenges for {6G} Wireless Systems: A Summary and Some Solutions}, 
  year={2022},
  volume={17},
  number={1},
  pages={16-26},
  doi={10.1109/MVT.2021.3136506}}

@ARTICLE{huang,
  author={Huang, Howard and Trivellato, Matteo and Hottinen, Ari and Shafi, Mansoor and Smith, Peter J. and Valenzuela, Reinaldo},
  journal={IEEE Transactions on Wireless Communications}, 
  title={Increasing downlink cellular throughput with limited network {MIMO} coordination}, 
  year={2009},
  volume={8},
  number={6},
  pages={2983-2989},
  doi={10.1109/TWC.2009.080179}}

@ARTICLE{qureshi,
  author={Forney, G. and Gallager, R. and Lang, G. and Longstaff, F. and Qureshi, S.},
  journal={IEEE Journal on Selected Areas in Communications}, 
  title={Efficient Modulation for Band-Limited Channels}, 
  year={1984},
  volume={2},
  number={5},
  pages={632-647},
  doi={10.1109/JSAC.1984.1146101}}

@ARTICLE{Forney,
  author={Forney, G.D. and Wei, L.-F.},
  journal={IEEE Journal on Selected Areas in Communications}, 
  title={Multidimensional constellations. {I.} Introduction, figures of merit, and generalized cross constellations}, 
  year={1989},
  volume={7},
  number={6},
  pages={877-892},
  doi={10.1109/49.29611}}

@ARTICLE{7307154,
  author={Böcherer, Georg and Steiner, Fabian and Schulte, Patrick},
  journal={IEEE Transactions on Communications}, 
  title={Bandwidth Efficient and Rate-Matched Low-Density Parity-Check Coded Modulation}, 
  year={2015},
  volume={63},
  number={12},
  pages={4651-4665},
  doi={10.1109/TCOMM.2015.2494016}}

@ARTICLE{pasupathy,
  author={Kschischang, F.R. and Pasupathy, S.},
  journal={IEEE Transactions on Information Theory}, 
  title={Optimal nonuniform signaling for {Gaussian} channels}, 
  year={1993},
  volume={39},
  number={3},
  pages={913-929},
  doi={10.1109/18.256499}}

@ARTICLE{cellfree,

  author={Ngo, Hien Quoc and Ashikhmin, Alexei and Yang, Hong and Larsson, Erik G. and Marzetta, Thomas L.},

  journal={IEEE Transactions on Wireless Communications}, 

  title={Cell-Free Massive {MIMO} Versus Small Cells}, 

  year={2017},

  volume={16},

  number={3},

  pages={1834-1850},

  doi={10.1109/TWC.2017.2655515}}

@INPROCEEDINGS{GLee,

  author={Lee, Gilwon and Rahman, Md Saifur and Onggosanusi, Eko},

  booktitle={2022 IEEE Wireless Communications and Networking Conference (WCNC)}, 

  title={{CSI} Feedback for Distributed {MIMO}}, 

  year={2022},

  volume={},

  number={},

  pages={2154-2159},

  doi={10.1109/WCNC51071.2022.9771853}}

@ARTICLE{1261332,
  author={Spencer, Q.H. and Swindlehurst, A.L. and Haardt, M.},
  journal={IEEE Transactions on Signal Processing}, 
  title={Zero-forcing methods for downlink spatial multiplexing in multiuser {MIMO} channels}, 
  year={2004},
  volume={52},
  number={2},
  pages={461-471},
  doi={10.1109/TSP.2003.821107}}

@ARTICLE{Richardson,
  author={Richardson, Tom and Kudekar, Shrinivas},
  journal={IEEE Communications Magazine}, 
  title={Design of Low-Density Parity Check Codes for 5G New Radio}, 
  year={2018},
  volume={56},
  number={3},
  pages={28-34},
  doi={10.1109/MCOM.2018.1700839}}

@ARTICLE{Richardson2,
  author={Richardson, T.J. and Shokrollahi, M.A. and Urbanke, R.L.},
  journal={IEEE Transactions on Information Theory}, 
  title={Design of capacity-approaching irregular low-density parity-check codes}, 
  year={2001},
  volume={47},
  number={2},
  pages={619-637},
  doi={10.1109/18.910578}}

@inproceedings{Jindal_2005,
   title={High {SNR} analysis of {MIMO} broadcast channels},
   url={http://dx.doi.org/10.1109/ISIT.2005.1523760},
   DOI={10.1109/isit.2005.1523760},
   booktitle={Proceedings. International Symposium on Information Theory, 2005. ISIT 2005.},
   publisher={IEEE},
   author={Jindal, N.},
   year={2005},
   pages={2310–2314} }

@ARTICLE{10054381,
  author={Wang, Cheng-Xiang and You, Xiaohu and Gao, Xiqi and Zhu, Xiuming and Li, Zixin and Zhang, Chuan and Wang, Haiming and Huang, Yongming and Chen, Yunfei and Haas, Harald and Thompson, John S. and Larsson, Erik G. and Renzo, Marco Di and Tong, Wen and Zhu, Peiying and Shen, Xuemin and Poor, H. Vincent and Hanzo, Lajos},
  journal={IEEE Communications Surveys \& Tutorials}, 
  title={On the Road to {6G}: Visions, Requirements, Key Technologies, and Testbeds}, 
  year={2023},
  volume={25},
  number={2},
  pages={905-974},
  doi={10.1109/COMST.2023.3249835}}

@ARTICLE{10634051,
  author={Na, Minsoo and Lee, Jaehyun and Choi, Giwan and Yu, Takki and Choi, Jeongsik and Lee, Jinyoung and Bahk, Saewoong},
  journal={IEEE Communications Magazine}, 
  title={Operator's Perspective on {6G: 6G Services}, Vision, and Spectrum}, 
  year={2024},
  volume={62},
  number={8},
  pages={178-184},
  doi={10.1109/MCOM.001.2400060}}

@ARTICLE{5741160,
  author={Dohler, M. and Heath, R.W. and Lozano, A. and Papadias, C.B. and Valenzuela, R.A.},
  journal={IEEE Communications Magazine}, 
  title={Is the {PHY} layer dead?}, 
  year={2011},
  volume={49},
  number={4},
  pages={159-165},
  doi={10.1109/MCOM.2011.5741160}}

@ARTICLE{Rusek,
  author={Rusek, Fredrik and Persson, Daniel and Lau, Buon Kiong and Larsson, Erik G. and Marzetta, Thomas L. and Edfors, Ove and Tufvesson, Fredrik},
  journal={IEEE Signal Processing Magazine}, 
  title={Scaling Up {MIMO}: Opportunities and Challenges with Very Large Arrays}, 
  year={2013},
  volume={30},
  number={1},
  pages={40-60},
  doi={10.1109/MSP.2011.2178495}}

@ARTICLE{Valenzuela,
  author={Karakayali, M.K. and Foschini, G.J. and Valenzuela, R.A.},
  journal={IEEE Wireless Communications}, 
  title={Network coordination for spectrally efficient communications in cellular systems}, 
  year={2006},
  volume={13},
  number={4},
  pages={56-61},
  doi={10.1109/MWC.2006.1678166}}

@ARTICLE{massivemimo,
  author={Larsson, Erik G. and Edfors, Ove and Tufvesson, Fredrik and Marzetta, Thomas L.},
  journal={IEEE Communications Magazine}, 
  title={Massive {MIMO} for next generation wireless systems}, 
  year={2014},
  volume={52},
  number={2},
  pages={186-195},
  doi={10.1109/MCOM.2014.6736761}}

@ARTICLE{rel18mimo,
  author={Jin, Huangping and Liu, Kunpeng and Zhang, Min and Zhang, Leiming and Lee, Gilwon and Farag, Emad N. and Zhu, Dalin and Onggosanusi, Eko and Shafi, Mansoor and Tataria, Harsh},
  journal={IEEE Journal on Selected Areas in Communications}, 
  title={Massive {MIMO} Evolution Toward {3GPP} Release 18}, 
  year={2023},
  volume={41},
  number={6},
  pages={1635-1654},
  doi={10.1109/JSAC.2023.3273768}}

@ARTICLE{foschini,
    author = {Foschini,G.and Gans,M.},
journal ={Wireless Personal Communications},
    title={On Limits of Wireless Communications in a Fading Environment when Using Multiple Antennas},
        year= {1998},
 number={6},
  pages={311-335},
}

@INPROCEEDINGS{gcs,
  author={Jia, Yinhua and Wu, Liangming and Xu, Changlong and Liu, Wei and Xu, Hao and Richardson, Tom},
  booktitle={2023 IEEE Globecom Workshops (GC Wkshps)}, 
  title={A Novel Design for Geometric Constellation Shaping}, 
  year={2023},
  volume={},
  number={},
  pages={1554-1559},
  doi={10.1109/GCWkshps58843.2023.10464443}}

@ARTICLE{shaping,
  author={Foschini, G. and Gitlin, R. and Weinstein, S.},
  journal={IEEE Transactions on Communications}, 
  title={Optimization of Two-Dimensional Signal Constellations in the Presence of {Gaussian} Noise}, 
  year={1974},
  volume={22},
  number={1},
  pages={28-38},
  doi={10.1109/TCOM.1974.1092061}}

@inproceedings{daoud2025pilot,
	title={Pilot Pattern Learning and Deep {CSI} Reconstruction for Efficient {MIMO} Systems},
	author={Daoud Burghal and Xinliang Zhang and Yan Xin and Younghan Nam},
	booktitle={Proc. IEEE Global Communications Conference (Globecomm)},
	pages={1--6},
	year={2025},
}

@article{burghal2023enhanced,
	title={{Enhanced AI-based CSI prediction solutions for massive {MIMO} in 5G and 6G systems}},
	author={Burghal, Daoud and Li, Yang and Madadi, Pranav and Hu, Yeqing and Jeon, Jeongho and Cho, Joonyoung and Molisch, Andreas F and Zhang, Jianzhong},
	journal={IEEE Access},
	volume={11},
	pages={117810--117825},
	year={2023},
}

@techreport{R1-2508800:sam,
	author = {Samsung},
	title = {{Design of 6GR air interface}},
	year = {2025},
	institution = "3GPP",
	number = {R1-2508800},
	address = {Dallas, USA},
	month = {November}
}

@article{o2017introduction,
  title={An introduction to deep learning for the physical layer},
  author={O'shea, Timothy and Hoydis, Jakob},
  journal={IEEE Transactions on Cognitive Communications and Networking},
  volume={3},
  number={4},
  pages={563--575},
  year={2017},
  publisher={IEEE}
}

@article{zappone2019wireless,
  title={Wireless networks design in the era of deep learning: Model-based, {AI}-based, or both?},
  author={Zappone, Alessio and Di Renzo, Marco and Debbah, M{\'e}rouane},
  journal={IEEE Transactions on Communications},
  volume={67},
  number={10},
  pages={7331--7376},
  year={2019},
  publisher={IEEE}
}

@article{lin2023embracing,
  title={{Embracing AI in 5G-advanced toward 6G: A joint 3GPP and O-RAN perspective}},
  author={Lin, Xingqin and Kundu, Lopamudra and Dick, Chris and Velayutham, Soma},
  journal={IEEE Communications Standards Magazine},
  volume={7},
  number={4},
  pages={76--83},
  year={2023},
  publisher={IEEE}
}

@article{hoydis2021toward,
  title={Toward a {6G AI}-native air interface},
  author={Hoydis, Jakob and Ait Aoudia, Fay{\c{c}}al and Valcarce, Alvaro and Viswanathan, Harish},
  journal={IEEE Communications Magazine},
  volume={59},
  number={5},
  pages={76--81},
  year={2021},
  publisher={IEEE}
}

@article{qin2019deep,
  title={Deep learning in physical layer communications},
  author={Qin, Zhijin and Ye, Hao and Li, Geoffrey Ye and Juang, Biing-Hwang Fred},
  journal={IEEE Wireless Communications},
  volume={26},
  number={2},
  pages={93--99},
  year={2019},
  publisher={IEEE}
}

@article{honkala2021deeprx,
  title={{DeepRx}: {Fully} convolutional deep learning receiver},
  author={Honkala, Mikko and Korpi, Dani and Huttunen, Janne MJ},
  journal={IEEE Transactions on Wireless Communications},
  volume={20},
  number={6},
  pages={3925--3940},
  year={2021},
  publisher={IEEE}
}

@inproceedings{cammerer2023neural,
  title={A neural receiver for {5G NR multi-user MIMO}},
  author={Cammerer, Sebastian and A{\"\i}t Aoudia, Fay{\c{c}}al and Hoydis, Jakob and Oeldemann, Andreas and Roessler, Andreas and Mayer, Timo and Keller, Alexander},
  booktitle={2023 IEEE Globecom Workshops (GC Wkshps)},
  pages={329--334},
  year={2023},
  organization={IEEE}
}

@article{ait2021end,
  title={End-to-end learning for {OFDM}: {From} neural receivers to pilotless communication},
  author={Ait Aoudia, Fay{\c{c}}al and Hoydis, Jakob},
  journal={IEEE Transactions on Wireless Communications},
  volume={21},
  number={2},
  pages={1049--1063},
  year={2021},
  publisher={IEEE}
}

@software{sionna,
 title = {Sionna},
 author = {Hoydis, Jakob and Cammerer, Sebastian and {Ait Aoudia}, Fayçal and
 Nimier-David, Merlin and Maggi, Lorenzo and Marcus, Guillermo and Vem, Avinash and Keller,
 Alexander},
 note = {https://nvlabs.github.io/sionna/},
 year = {2022},
 version = {1.2.1}
}

@ARTICLE{kundu2025ai,
  author={Kundu, Lopamudra and Lin, Xingqin and Gadiyar, Rajesh and Lacasse, Jean-Francois and Chowdhury, Shuvo},
  journal={IEEE Communications Magazine}, 
  title={{AI-RAN}: Transforming {RAN} with {AI}-Driven Computing Infrastructure}, 
  year={2026},
  volume={64},
  number={1},
  pages={168-174},
}

@ARTICLE{Giuse1998BICM,
  author={Giuseppe Caire and Giorgio Taricco and Ezio Biglieri},
  journal={IEEE Transactions on Information Theory}, 
  title={Bit-interleaved coded modulation}, 
  year={1998},
  month = {May},
  volume={44},
  number={3},
  pages={927-946},
}

@ARTICLE{Steph2003MOD,
  author={St\'{e}phane Y. Le Goff},
  journal={IEEE Transactions on Information Theory}, 
  title={Signal constellations for bit-interleaved coded modula
tion}, 
  year={2003},
  month = {Jan.},
  volume={49},
  number={1},
  pages={307-313},
}

@ARTICLE{Jon2017MOD,
  author={Jon, Barrueco and Jon, Montalban and Cristina, Regueiro and Manuel, Velez and Juan Luis Ordiales and Heung-Mook Kim  and Sung-Ik Park and Sunhyoung Kwon},
  journal={IEEE Transactions on Broadcasting}, 
  title={Constellation design for bit-interleaved coded
modulation ({BICM}) systems in advanced broadcast standards}, 
  year={2017},
  month = {Dec.},
  volume={63},
  number={4},
  pages={603-614},
}

@standard{ATSC2017MOD,
  title         = "Physical Layer Protocol",
  howpublished  = "document A/322:2017",
  organization  = "Advanced Television System Committee",
  address       = "Washington, D.C.",
  type          = "",
  number        = "",
  month         = {June},
  year          = "2017",
  url           = ""
}

@techreport{sonygcs,
    author = {Sony},
    title = {Performance of Non-Uniform Constellations in {NR}},
    institution = {{3GPP} RAN1\#87},
    year = 2026,
    month = {Nov.},
    number = {{R1-1611543}},
    address = {Reno, USA}}

@article{Qualcoding2026,
	title={Channel Coding for {6GR}},
	author={Qualcomm},
	journal = {3GPP R1-2603009, 3GPP TSG RAN WG1},
	month={April},
	year={2026},
    address = {St. Julian's, Malta,},
}

@article{Qualc2602,
	title={Channel Coding for {6GR}},
	author={Qualcomm},
	journal = {3GPP R1-2601269, 3GPP TSG RAN WG1},
	month={Feb.},
	year={2026},
    address = {Gothenburg, Sweden},
}

@inproceedings{ts901,
	title={{5G}; Study on channel model for frequencies from 0.5 to 100 {GHz}},
	author={3GPP},
	booktitle={TS 38.901, version 19.1.0},
	pages={97},
	year={Oct. 2025},
}

@techreport{ref1,
  author       = "{Moderator (Nokia)}",
  title        = "Feature Lead Summary \#5 on {6G} Waveform",
  institution  = "{3GPP}",
  year         = "2026",
  month        = May,
number       = "{{R1-2603729}}",  
address = {Dalian, China}
}

@misc{ref2,
  author       = "Imran Ali Khan and Saif Khan Mohammed and Ronny Hadani and Ananthanarayanan Chockalingam and Robert Calderbank and Anton Monk and Shachar Kons and Shlomo Rakib and Yoav Hebron",
  title        = "Waveform for Next Generation Communication Systems: Comparing {Zak-OTFS with OFDM}",
  year         = "2025",
  month        = may,
  eprint       = "2505.13966",
  archivePrefix = "arXiv",
  primaryClass = "eess.SP",
  note         = "arXiv preprint"
}

@article{ref6,
  author       = "S. R. Mattu and N. Mehrotra and S. K. Mohammed and others",
  title        = "Low-Complexity Equalization of {Zak-OTFS} in the Frequency Domain",
  journal      = "npj Wireless Technology",
  volume       = "2",
  number       = "9",
  year         = "2026",
  doi          = "10.1038/s44459-025-00011-0"
}

@techreport{ref7,
  author       = "{Samsung}",
  title        = "Discussion on Waveform and Frame Structure",
  institution  = "3GPP",
  number       = "R1-2604202",
  address      = "Dalian, China",
  month        = may,
  year         = "2026",
  note         = "{RAN1} \#125, May 18--21, 2026"
}

@techreport{ref8,
  author       = "{Qualcomm}",
  title        = "{Waveforms for {6GR}}",
  institution  = "3GPP",
  number       = "R1-2604689",
  month        = may,
  year         = "2026",
  note         = "3GPP TSG {RAN} WG1 \#125"
}

@techreport{3gpp38914,
  author       = "{3GPP}",
  title        = "{Study on 6G Scenarios and requirements}",
  institution  = "3GPP",
  number       = "TR 38.914 V20.0.0",
  month        = Jul,
  year         = "2026",
  note         = ""
}

@techreport{3gpp387601,
  author       = "{3GPP}",
  title        = "{Study on 6G Radio RAN1 aspects}",
  institution  = "3GPP",
  number       = "TR 38.760-1 V20.0.0",
  month        = Sep,
  year         = "2026",
  note         = ""
}

@techreport{ref10,
  author       = "{Samsung}",
  title        = "{Discussion on DL-based CSI acquisition and CSI-RS design}",
  institution  = "3GPP",
  number       = "TDoc R1-2605065",
  month        = may,
  year         = "2026",
  note         = "3GPP TSG RAN WG1 Meeting 125"
}

@ARTICLE{9954153,

  author={Xu, Jialong and Ai, Bo and Wang, Ning and Chen, Wei},

  journal={IEEE Journal on Selected Areas in Communications}, 

  title={Deep Joint Source-Channel Coding for {CSI} Feedback: An End-to-End Approach}, 

  year={2023},

  volume={41},

  number={1},

  pages={260-273},

  doi={10.1109/JSAC.2022.3221963}}

@ARTICLE{10660530,
  author={Guo, Yiran and Chen, Wei and Xu, Jialong and Li, Lun and Ai, Bo},
  journal={IEEE Transactions on Vehicular Technology}, 
  title={Deep Joint {CSI} Feedback and Multiuser Precoding for {MIMO OFDM} Systems}, 
  year={2025},
  volume={74},
  number={1},
  pages={1730-1735},
  doi={10.1109/TVT.2024.3452409}}

@ARTICLE{11563906,
  author={Jiang, Nan and Kim, Younsun and Yang, Bowen and Qi, Hang and Xiao, Yu and Cui, Yupeng and Sun, Feifei and Qian, Chen and Yu, Bin and Sun, Chengjun and Ji, Hyoung-ju and Jang, Min},
  journal={IEEE Transactions on Vehicular Technology}, 
  title={{CSI} Feedback via Joint Source-Channel Coding and Modulation: A Comprehensive Study for {6G} Realization}, 
  year={2026},
  volume={},
  number={},
  pages={1-17},
  doi={10.1109/TVT.2026.3703766}}

@online{sun2025csi,
  author    = {Sun, Feifei and Abebe, Ameha Tsegaye and Ji, Hyoungju and Onggosanusi, Eko},
  title     = {6{G} {AI}/{ML} for Physical-layer: Part {II} - {CSI} Compression with {JSCM}},
  journal   = {Samsung Research Blog},
  date      = {2025-12-30},
  url       = { https://research.samsung.com/blog/6G-AI-ML-for-Physical-layer-Part-II-CSI-Compression-with-JSCM}
}

@article{delfeld2025sparse,
  title={Sparse {MIMO-OFDM} channel estimation via {RKHS} regularization},
  author={Delfeld, James and Marti, Gian and Dick, Chris},
  journal={arXiv preprint arXiv:2511.20082},
  year={2025}
}

@ARTICLE{5595728,
  author={Marzetta, Thomas L.},
  journal={IEEE Transactions on Wireless Communications}, 
  title={Noncooperative Cellular Wireless with Unlimited Numbers of Base Station Antennas}, 
  year={2010},
  volume={9},
  number={11},
  pages={3590-3600},
  doi={10.1109/TWC.2010.092810.091092}}

@ARTICLE{11186253,
  author={Lin, Xingqin},
  journal={IEEE Communications Standards Magazine}, 
  title={A Tale of Two Mobile Generations: {5G-Advanced and 6G} in {3GPP} Release 20}, 
  year={2025},
  volume={},
  number={},
  pages={1-9},
  doi={10.1109/MCOMSTD.2025.3613219}}

%\begin{IEEEbiography}{Mansoor Shafi}
%Biography text here.
%\end{IEEEbiography}

\end{document}